# Mechanized Noninterference for Gradual Security

TIANYU CHEN and JEREMY G. SIEK, Indiana University, USA

This paper presents the first machine-checked proof of noninterference for a language with gradual information-flow control, thereby establishing a rock solid foundation for secure programming languages that give programmers the choice between runtime versus compile-time enforcement. Along the way we uncovered a flaw in one of the noninterference proofs in the literature, and give a counterexample for one of the main lemmas. The particular language studied in this paper, $\lambda^{\star}_{\mathsf{SEC}}$, is based on the GLIO language of Azevedo de Amorim et al. [2020]. To make the design more accessible to other researchers, this paper contributes the first traditional semantics for the language, that is, we define compilation from $\lambda^{\star}_{\mathsf{SEC}}$ to a cast calculus and design a reduction semantics for the latter that includes blame tracking. In addition to the proof of noninterference, we also mechanize proofs of type safety, determinism, and that compilation preserves types.

## 1 INTRODUCTION

Information-flow control (IFC) ensures that information transfers within a program adhere to a security policy, for example, by preventing high-security data from flowing to a low-security channel. This adherence can be enforced statically using a type system [Myers 1999; Myers and Liskov 1997; Volpano et al. 1996], or dynamically using runtime monitoring [Askarov and Sabelfeld 2009; Austin and Flanagan 2009; Austin et al. 2017; Devriese and Piessens 2010; Stefan et al. 2011], or with a combination of the two [Chandra and Franz 2007; Shroff et al. 2007; Zheng and Myers 2005]. The two approaches have complementary strengths and weaknesses; the dynamic approach requires less effort from the programmer while the static approach provides stronger guarantees and less runtime overhead.

Taking inspiration from gradual typing [Siek and Taha 2006, 2007], researchers have explored how to give programmers control over which parts of the program are secured statically versus dynamically. The main challenge in such systems is controlling the flow of values (and information) between the static and dynamic regions of code, which is traditionally accomplished using runtime casts. Disney and Flanagan [2011] design a cast calculus with IFC for a pure lambda calculus and prove noninterference. Fennell and Thiemann [2013] design a cast calculus for an imperative, object-oriented language [Fennell and Thiemann 2015], using the no-sensitive-upgrade runtime checks of Austin and Flanagan [2009].

Toro et al. [2018] analyze the semantics of runtime casts through the lense of Abstracting Gradual Typing [Garcia et al. 2016], and observe that security typing should induce "free theorems" [Wadler 1989] about noninterference, but that prior cast calculi do not. Toro et al. [2018] propose a new semantics for casts with the $\text{GSL}_{\text{Ref}}$ calculus and prove noninterference. However, they also discover that there is a tension between gradual security and the gradual guarantee, an important property of gradual typed languages [Siek et al. 2015]. Azevedo de Amorim et al. [2020] pinpoint one source of the tension: the type-guided classification performed by casts in $\text{GSL}_{\text{Ref}}$. They propose a new gradually typed source language, GLIO, give it a denotational semantics, and prove that it satisfies both noninterference and the gradual guarantee. Bichhawat et al. [2021] locate another source of the tension, and instead resolve it via a hybrid approach that leverages static analysis to determine the write effects in untaken branches.

Meanwhile, advances in proof assistants [Bove et al. 2009; Nipkow et al. 2007; The Coq Dev. Team 2004] have made it feasible to produce machine-checked (aka. mechanized) proofs of metatheoretic properties of programming languages [Aydemir et al. 2005]. Given the sensitive nature of information-flow control, there is greater desire to know that the proofs are correct. Indeed, Stefan et al. [2017] develop a mechanized proof of noninterference for LIO, a functional language with dynamic IFC. Xiang and Chong [2021] take this a step further and produce a mechanized proof on noninterference for an imperative object-oriented language.

The primary contribution of this paper is a mechanized proof of noninterference for a gradual security-typed language named $\lambda^{\star}_{\text{SEC}}$ that is similar GLIO [Azevedo de Amorim et al. 2020]. The secondary contribution is the definition of $\lambda^{\star}_{\text{SEC}}$ via traditional means, that is, through a cast calculus and reduction semantics, to make the semantics accessible to more researchers.

When a $\lambda^{\star}_{\text{SEC}}$ program is fully statically typed, the type system enforces information flow security just like that of a static security-typed language. Unlike a static language, in $\lambda^{\star}_{\text{SEC}}$, programmers do not have to supply all the static type information up-front when developing the software. They may instead opt for less precise type annotations by using the unknown security label, written $\star$, which defers some of the IFC checks until runtime. During program execution, security labels are attached to values and the deferred IFC checks inspect those labels to guarantee secure information flow. This approach alleviates some of the pain of the programmer wrestling with the type checker, while keeping the security level of data unambiguous.

This paper makes the following technical contributions:

- First mechanized proof of noninterference for a gradual information-flow language (§ 6.3).
- Design of a cast calculus for gradual information-flow, including blame tracking (§ 5).
- Mechanized proofs of type safety (§ 6.1), determinism for the cast calculus under erasure (§ 6.2), and that compilation preserves types (§ 6.4).
- Counterexample to a noninterference theorem of Fennell and Thiemann [2013] (§ 3.1).

The semantics of $\lambda^{\star}_{\text{SEC}}$ and its cast calculus, and all the above-mentioned proofs are mechanized in the Agda proof assistant and are available at the following URL:

https://github.com/Gradual-Typing/LambdaSecStar/archive/refs/tags/v0.9-alpha.tar.gz

## 2 EXAMPLE PROGRAMS: $\lambda^{\star}_{\text{SEC}}$ IN ACTION

In this section we present example programs so the reader can get a taste of $\lambda^{\star}_{\text{SEC}}$ and establish the intuition that $\lambda^{\star}_{\text{SEC}}$ satisfies noninterference. We briefly review the basics of IFC and gradual typing in Section 2.1. We then compare $\lambda^{\star}_{\text{SEC}}$ with $\text{GSL}_{\text{Ref}}$ and GLIO with respect to the dynamic gradual guarantee in Section 2.2.

For simplicity, we use the security lattice $\langle\{\texttt{high}, \texttt{low}\}, \preccurlyeq, \curlyvee, \curlywedge\rangle$, where `high` is for private data while `low` for publicly disclosable data. They satisfy $\texttt{low} \preccurlyeq \texttt{high}$ and $\texttt{high} \not\preccurlyeq \texttt{low}$, meaning that information can flow from public sources to private sinks but not the other way around. Types have security labels associated with them, for example, $\texttt{Bool}_{\texttt{high}}$ is the type for booleans with high security and $\texttt{Unit}_{\texttt{low}}$ is the type for the unit value with low security. We use () as a shorthand for the value of $\texttt{Unit}_{\texttt{low}}$. We model I/O with two functions, `user-input` and `publish`: the former returns a high-security boolean that represents sensitive input information; the latter takes a low-security boolean and publishes it into a publicly visible channel. They have the following signatures:

```
user-input : Unit_low → Bool_high
publish    : Bool_low → Unit_low
```

## 2.1 Basics of Gradual Information Flow Security

Consider a program that takes in high-security user input and publishes the return value of `fconst`:

```
let fconst = (λ b : Bool_high. false_low )_low in
let input  = user-input () in
let result = fconst input in
  publish result
```

The program is fully statically typed, as there are no uses of the unknown label. The program type-checks and runs without error. Indeed, a malicious party cannot infer anything about high-security input, because 1) the return value of `fconst` is always the same value $\texttt{false}_{\texttt{low}}$ 2) the value $\texttt{false}_{\texttt{low}}$ is of low security, so the explicit flow into `publish` is allowed.

If we replace `fconst` with the identity function on $\texttt{Bool}_{\texttt{low}}$, `fid`, the program becomes ill-typed, because our type system does not allow the explicit flow from the high-security input to `fid`:

```
let fid    = (λ b : Bool_low . b)_low in
let input  = user-input () in
let result = fid input in  // error, input is high security but fid expects low
  publish result
```

Sometimes the observable behaviors of a program can depend on its branching structure. If some of the branch conditions have a data dependency on high-security input, a malicious party might be able to infer it from the observable behaviors, giving rise to illegal *implicit flows* [Denning 1976], which must be ruled out in order to guarantee security.

Consider the following program in which the function `flip` contains one if-expression, whose branch condition is dependent on high-security user input. Its two branches return different low-security booleans, creating a potential implicit flow from high to low:

```
let flip : Bool_high → Bool_low = (λ b : Bool_high. if b then false_low else true_low )_low in
let input  = user-input () in
let result = flip input in
  publish result
```

This program is rejected by the type checker, thereby preventing an information leak through an implicit flow. The programmer annotates the return type of `flip` thinking that it must return $\texttt{Bool}_{\texttt{low}}$, because both branches contain low-security values. However, because the branch condition is of high security the type of the if-expression as a whole must be $\texttt{Bool}_{\texttt{high}}$. In particular, the type checker computes the security level of a conditional to be the join of its branches (both `low`) and the branch condition (`high`), $\texttt{low} \curlyvee \texttt{high} = \texttt{high}$. The `flip` function is expected to return $\texttt{Bool}_{\texttt{low}}$ according to its type annotation, but returns $\texttt{Bool}_{\texttt{high}}$ because of the conditional, $\texttt{high} \not\preccurlyeq \texttt{low}$, so the program is ill-typed.

To summarize, $\lambda^{\star}_{\texttt{SEC}}$ behaves just like a static security-typed language in the above examples. When everything is statically typed, the type system of $\lambda^{\star}_{\texttt{SEC}}$ guards against illegal information flows, whether explicit or implicit. Meanwhile, in addition to *concrete* security labels, `low` and `high`, $\lambda^{\star}_{\mathtt{SEC}}$ also provides another label $\star$, which stands for statically *unknown* security level. We explain how the unknown security level works in the next paragraph.

To repair the `flip` example, the programmer has two choices. They can either invest time and effort in reasoning rigorously about the program and providing precise and correct type information, or instead change the type annotations to be more dynamic. Suppose they chooses the latter approach, changing the argument annotation on the $\lambda$ from $\mathtt{Bool}_{\mathtt{high}}$ to $\mathtt{Bool}_{\star}$ and changing its type signature annotation accordingly. In the meantime, the return type remains $\mathtt{Bool}_{\mathtt{low}}$, to conform with the signature of `publish`. Line 1 thus becomes:

```
let flip : Bool_⋆ → Bool_low = (λ b : Bool_⋆ . if b then false_low else true_low)_low in
```

This change makes the program well-typed. The IFC enforcement of the implicit flow is deferred until runtime, because the branch condition now has type $\mathtt{Bool}_{\star}$, with an unknown security level.

The dynamic semantics of $\lambda^{\star}_{\mathtt{SEC}}$ is defined by compilation into $\lambda^{\Rightarrow}_{\mathtt{SEC}}$ by inserting casts. A *cast calculus* is an intermediate representation where all casts are made explicit. We define $\lambda^{\Rightarrow}_{\mathtt{SEC}}$ and present the compilation rules formally in Section 5. The general idea is to expose a cast wherever an implicit cast occured in the typing derivation of the $\lambda^{\star}_{\mathtt{SEC}}$ term. The result of cast insertion on this program is the following $\lambda^{\Rightarrow}_{\mathtt{SEC}}$ term:

```
let flip   = (λ b. (if b then false_low else true_low) ⟨Bool_⋆ ⇒^p Bool_low⟩ )_low in
let input  = user-input () in
let result = flip (input ⟨Bool_high ⇒^q Bool_⋆⟩ ) in
  publish result
```

where two casts are made explicit. Each cast has a *blame label* attached to it. In case a cast fails, it produces a cast error, called *blame*, that contains its label. In this way, the programmer knows which cast is causing the problem. This feature is often referred to as *blame tracking* [Findler and Felleisen 2002; Wadler and Findler 2009].

The first cast, which has blame label $p$, casts the result of the if-expression to a low security Boolean. We refer to such casts from $\star$ to a concrete label as *projections*. The second cast with blame label $q$ casts input from $\mathtt{Bool}_{\mathtt{high}}$ to $\mathtt{Bool}_{\star}$, to conform with the parameter type of `flip`. We call the casts from a concrete label to $\star$ as *injections*.

If we run the program with $\mathtt{true}_{\mathtt{high}}$ or $\mathtt{false}_{\mathtt{high}}$ as input, the $\lambda^{\Rightarrow}_{\mathtt{SEC}}$ term reduces to `blame` $p$ in either situation. The illegal implicit flow is captured by the runtime. Regardless of the branch taken, the observable behavior is always the same, so no information is leaked. The following shows the highlights of the reduction to blame with input $\mathtt{true}_{\mathtt{high}}$, which we discuss in the following paragraph.

```
⟶*  let result = ((λ b. (if b then false_low else true_low) ⟨Bool_⋆ ⇒^p Bool_low⟩)_low
                  (true_high ⟨Bool_high ⇒^q Bool_⋆⟩)) in ...                          (1)

⟶*  let result = prot low ((if (true_high ⟨Bool_high ⇒^q Bool_⋆⟩) then false_low ...)
                  ⟨Bool_⋆ ⇒^p Bool_low⟩) in ...                                       (2)

⟶*  let result = prot low ((prot high false_low) ⟨Bool_high ⇒^q Bool_⋆⟩
                  ⟨Bool_⋆ ⇒^p Bool_low⟩) in ...                                       (3)

⟶*  let result = prot low (false_high ⟨Bool_high ⇒^q Bool_⋆⟩ ⟨Bool_⋆ ⇒^p Bool_low⟩) in ...   (4)

⟶*  blame p                                                                            (5)
```

The reduction sequence begins by evaluating the first two `let`s by substituting `flip` with the lambda and `input` with $\mathtt{true}_{\mathtt{high}}$ (1). The next step is function application, which substitutes `b` with $\mathtt{true}_{\mathtt{high}}$ injected from $\mathtt{Bool}_{\mathtt{high}}$ to $\mathtt{Bool}_{\star}$ and encloses the body of the function in a protection term `prot low` because the lambda itself was of low security (2). (We say more about protection terms shortly.) The next step is to reduce the `if` conditional, which gives rise to the implicit flow of interest. The condition value is a `true` of `high` security, so the `if` reduces to the then-branch surrounded by a high security protection term and a cast from $\mathtt{Bool}_{\mathtt{high}}$ to $\mathtt{Bool}_{\star}$ (3). As is standard for security-typed languages [Fennell and Thiemann 2013; Heintze and Riecke 1998; Toro et al. 2018], the protection term ensures that the computed value and the side effects of its sub-term must be at least as secure as the security level of the protection term. In this case the protection term turns $\mathtt{false}_{\mathtt{low}}$ into $\mathtt{false}_{\mathtt{high}}$ (4). Next the sequence of two casts, from $\mathtt{Bool}_{\mathtt{high}}$ to $\mathtt{Bool}_{\mathtt{low}}$, trigger a runtime error because a high security value is not allowed to be cast to low security. Following standard practice, the blame goes to the projecting cast, so label $p$ is blamed (5).

## 2.2 Mutable References and Graduality

In $\lambda^{\star}_{\mathsf{SEC}}$ we use the *no-sensitive-upgrade* (NSU) [Austin and Flanagan 2009] technique to protect against illegal implicit flows through the heap. An NSU check happens at runtime when there is insufficient information to determine statically whether a heap write operation is secure or not. Consider the following well-typed program in $\lambda^{\star}_{\mathsf{SEC}}$:

```
let input : Bool_⋆ = user-input () in
let a     = ref low true_low in
let _     = if input then a := false_low else a := true_low in
  publish (! a)
```

The assignments in the two branches try to write different low-security booleans to address $a$, depending on a branch condition whose security level is statically unknown because of the type annotation $\star$. If the branch condition turns out to be high security, and if the assignments were successfully, the program would leak information via an implicit flow. Fortunately, if we run this program, it reduces to an NSU error regardless of the input, thanks to the NSU technique. The way NSU checking works is that a security label is associated with the current program counter and then at the point of every assignment, the system compares the program counter's security label PC with the security level of the memory location, making sure that the later is at least as high as the former. In the above example, the NSU check fails because the program counter's label is high during the execution of the branch but the write is to low memory. In general, all memory locations allocated or mutated must have security levels that are higher than the program counter's label. In this paper, we refer to the program counter's label as the *dynamic PC* and the type system's approximation of it as the *static PC*.

The dynamic heap policy of $\mathrm{GSL}_{\mathsf{Ref}}$ [Toro et al. 2018] is also based on NSU checks. Interestingly, the authors of $\mathrm{GSL}_{\mathsf{Ref}}$ claim that there is a tension between NSU and graduality. Consider the following pair of programs adapted from Section 6.3:

[**Left:** more precise, more static]

```
let x = user-input () in
let y = ref Bool_high true_high in
  if x then (y := false_high) else ()
```

[**Right:** less precise, more dynamic]

```
let x = user-input () in
let y = ref Bool_⋆ true_⋆ in
  if x then (y := false_high) else ()
```

The dynamic gradual guarantee (DGG) says that when moving type annotations to be less precise, the runtime behaviors of a program remain the same. In the above example, both variants type check but evaluate to different results, thus violating the DGG. Let us examine their runtime behaviors in further detail. The fully static program on the left runs without error regardless of the input being $\texttt{true}_{\texttt{high}}$ or $\texttt{false}_{\texttt{high}}$. Based on NSU, $\text{GSL}_{\text{Ref}}$'s heap policy allows assignments where the security effect subsumes the lower bound of the possible security levels that the memory location can have. At assignment, we know that y references a high-security memory cell and PC is high. The assignment on line 3 is allowed, because $\texttt{high} \preccurlyeq \texttt{high}$. We change the type annotations to be less precise by replacing some highs with ⋆ and get the program on the right. When we run the program with input $\texttt{true}_{\texttt{high}}$, the assignment will be conservatively rejected by the NSU check. This is because $\text{GSL}_{\text{Ref}}$ considers ⋆ corresponding to the interval [low, high], the lower bound of which is not subsumed by a high PC. Therefore, the runtime behaviors on the same input differ: the more precise program (left) runs successfully while the less precise one (right) errors.

GLIO is proved to satisfy DGG by its authors [Azevedo de Amorim et al. 2020]. Consider the loose translation of the example above into GLIO:

```
1  f :: Lab high Bool → LIO Unit
2  f x = do
3    b :: Lab high Bool ← toLab high true
4    b' ← unlabel b
5    y  ← new high b'
6    x' ← unlabel x
7    if x' then set y false
8          else return unit
9
10 do { in ← input; f in }
```

```
1  f :: Lab ⋆ Bool → LIO Unit
2  f x = do
3    b :: Lab ⋆ Bool ← toLab high true
4    b' ← unlabel b
5    y  ← new high b'
6    x' ← unlabel x
7    if x' then set y false
8          else return unit
9
10 do { in ← input; f in }
```

NSU checks pass and executions are successful for both programs. But there are two major differences from $\text{GSL}_{\text{Ref}}$: 1) only the labels on type annotations become ⋆ when migrating to dynamic, while labels on values and new memory locations remain concrete (i.e., not ⋆) (line 3 and 5); 2) there is no "type-guided classification" of data, in other words, casts only check for compatibility between types but never modify the labels on values. These design choices enable GLIO to reconcile the use of NSU and the DGG.

We adopt the two design choices of GLIO in $\lambda^{\star}_{\texttt{SEC}}$. The example becomes:

```
1  let x = user-input () in
2  let y = ref high true_high in
3    if x then (y := false_high) else ()
```

```
1  let x = user-input () in
2  let y = ref high (true_high : Bool_⋆) in
3    if x then (y := false_high) else ()
```

Similar to GLIO, both variants reduce to the unit value regardless of the input, thereby not violating DGG. When the program moves toward dynamic, an ⋆ annotation is added, while the label on the boolean constant and the label of the new memory location remain concrete (right, line 2), similar to GLIO. In other words, only labels *on types* are allowed to decrease in precision; labels on objects (values, memory locations ...) shall always be concrete. Also similar to GLIO, our system ditches type-guided classification, for example: $\texttt{true}_{\texttt{low}} \langle \texttt{Bool}_{\texttt{low}} \Rightarrow \texttt{Bool}_{\star} \rangle \langle \texttt{Bool}_{\star} \Rightarrow \texttt{Bool}_{\texttt{high}} \rangle \longrightarrow \texttt{true}_{\texttt{low}}$. Type annotations are compiled into explicit casts and casts never modify labels on the values that represent data (not "classifying data"). We will elaborate on the design of $\lambda^{\star}_{\texttt{SEC}}$ in Section 4.

## 3 DESIGN OF THE MECHANIZED NONINTERFERENCE PROOF

The complexity of a mechanized proof can vary considerably depending on the technical choices regarding the definition of the semantics and the proof strategy. In this section we give a high level discussion of our choices when designing the noninterference proof in Agda. We adopt the usual statement of termination-insensitive noninterference as the security guarantee of $\lambda^{\star}_{\mathtt{SEC}}$: a potentially malicious observer cannot discover the secretive inputs based on the computation results (values) produced by multiple successful executions of a $\lambda^{\star}_{\mathtt{SEC}}$ program.

We choose the erasure approach [Fennell and Thiemann 2013; Li and Zdancewic 2010; Stefan et al. 2017, 2011, 2012] as our proof technique. The basic idea is that for a low-privilege observer, high-security parts of a program that cannot be seen do not matter and can be "erased" to a single opaque value ●. In this way, all secretive inputs erase to ●, so a program substituted with different inputs always erases to the same term. Noninterference is a straightforward corollary of 1) simulation between the original program and the erased program and 2) the erased program evaluates deterministically. We define the dynamics of $\lambda^{\Rightarrow}_{\mathtt{SEC}}$ terms using big-step operational semantics that is straightforwardly derived from our small-step semantics (Section 5.4). We give an overview of our proof of noninterference in Section 3.2.

Alternatively, we could have based our mechanization on some of the pen-and-paper proofs of other gradual security languages. For example, Toro et al. [2018], use step-indexed logical relations, but there is no support for that approach in Agda, and building it would be a complex undertaking. Another approach is to define a denotational semantics and prove noninterference by relating the denotations [Azevedo de Amorim et al. 2020]. But again, building the infrastructure for such denotational semantics in Agda would require a major up-front investment. Our proof and that of Fennell and Thiemann [2013] both apply the erasure technique, however their simulation lemma is stated using small-step semantics and has a proof-breaking flaw, which we describe in Section 3.1.

### 3.1 Counterexample to Fennell and Thiemann [2013]

Fennell and Thiemann [2013] present a cast calculus named ML-GS and claim that it satisfies noninterference. Their small-step semantics takes the form $M \mid \mu \mid pc \longrightarrow M' \mid \mu'$, in which a term $M$ reduces to $M'$ while changing heap $\mu$ into $\mu'$, and $pc$ is the current PC of the computation.

Their proof depends on a simulation lemma (Lemma 2) between ML-GS and ML-GS$_L$. The latter is extended with an "opaque" value that all high-security parts of the program are erased to. We use ● the for opaque value and $\epsilon$ for the erasure function.

Lemma 2 (Fennell and Thiemann [2013]) *If* $M \mid \mu \mid \mathtt{low} \longrightarrow M' \mid \mu'$*, then* $\epsilon M \mid \epsilon\mu \mid \mathtt{low} \longrightarrow^{*} \epsilon M' \mid \epsilon\mu'$*.*

Consider creating a reference of a boolean of high security where the reference itself is low security, which then takes the following reduction step:

$$\mathtt{new}^{\mathtt{low}}\ \mathtt{true}^{\mathtt{high}} \mid \mu \mid \mathtt{low} \longrightarrow a^{\mathtt{low}} \mid \mu' \quad \text{, where } \mu' = a \mapsto \mathtt{true}^{\mathtt{high}} :: \mu \tag{6}$$

These terms erase as follows

$$\epsilon(\mathtt{new}^{\mathtt{low}}\ \mathtt{true}^{\mathtt{high}}) = \mathtt{new}^{\mathtt{low}}\ \bullet \qquad \text{and} \qquad \epsilon(a^{\mathtt{low}}) = a^{\mathtt{low}}$$

but $\mathtt{new}^{\mathtt{low}}\ \bullet$ does not reduce to $a^{\mathtt{low}}$. Instead it reduces to ●.

### 3.2 Overview of Our Noninterference Proof

Let us see if there is a straightforward fix to the counterexample in Section 3.1. Perhaps we could have a reduction rule that goes from $\mathtt{new}^{\mathtt{low}}\ \bullet$ to some address $a$ that is in sync with the unerased side. However, it is difficult to choose which address $a$ to allocate. When we erase $\mu$ to $\epsilon\mu$, all locations that store high-security values are erased and we end up with fewer heap cells. If we naïvely choose a fresh address in $\epsilon\mu$ it may be one that is already in use in $\mu$ for a different allocation, making it difficult to synchronize the heap $\mu$ with the heap on the erased side. This motivates us to rethink the heap model and revise the erasure function.

**Heap model.** We employ a split heap model that indexes low-security and high-security cells separately. A memory address $a = n_{\hat{\ell}}$, where $\hat{\ell}$ indicates whether it points to the high-security half or the low-security half and $n$ is the index of the cell in the half-heap. Each half-heap is represented in Agda as an association list that maps addresses to values. The high-heap can store low-security values, but the low-heap cannot store high-security values. When a reference is created, the programmer needs to explicit specify whether the new memory location is low-security or high-security, as in the examples of Section 2.2. In Agda we cannot hand-wave regarding the address being fresh, so we specifically choose the current length of the half-heap as the index part of our new address. The memory operations to the low-heap on the erased side mirror those on the unerased side, so the addresses are synchronized. When assignment happens, we know precisely which half-heap the address is referencing from its $\hat{\ell}$. Leveraging NSU checking, we can prove a lemma that all side effects that happen under a high PC only affect the high half of the heap, so they are not observable at the low privilege level.

**Erasure.** The intuition is that we erase everything that a low-privilege observer cannot see. This includes high-security constants, functions, and addresses. The erasure of address terms require some extra care: not only are address terms that are high-security themselves erased to ●, but also those addresses that point to the high-heap. In other words, only address terms shaped $(\texttt{addr}\ a)_{\texttt{low}}$ where $a = n_{\texttt{low}}$ (both being low) are not erased. We erase the terms related to gradual typing, specifically, the cast terms and PC-cast terms by discarding those casts and recursively erasing their sub-terms. As for the heap, erasure discards the high-half. For the low-half, we retain all the heap cells but apply the erasure function to their contents.

**Big-step semantics.** We formulate noninterference using a big-step semantics. The reason that we prefer big-step is that erasure-based proofs of noninterference rely on determinism of evaluation of the erased term, but that is difficult to achieve in a small-step semantics while also establishing a simulation between the original and erased program. For example, one cannot decide whether an NSU check on an opaque term $● \ \texttt{:=}^{?}\ M$ should succeed or not because we can't access the security level of the memory location corresponding to the erased address. For the purposes of establishing the simulation, one might consider defining two reduction rules, one that results in a checked term while the other fails with NSU error, but that would give rise to non-determinism.

Our big-step semantics is straightforwardly derived from the small-step semantics, but simplified because it relates terms to values, while leaving out rules that generate or propagate errors. This is because the theorem statement of termination-insensitive noninterference only concerns successful executions that produce values. In our noninterference proof (Section 6.3), we correct "Lemma 2" into Lemma 7 and prove the revised lemma.

## 4 $\lambda^{\star}_{\texttt{SEC}}$: THE SURFACE LANGUAGE

In this section we present the formal definition of $\lambda^{\star}_{\texttt{SEC}}$. Our high level design goal is to create a surface language whose meta-theory is easy to reason about in a mechanized way. Rather than being creative about individual language features, $\lambda^{\star}_{\texttt{SEC}}$ is more about rearranging and recombining the design choices in existing gradual security languages, such as $\text{GSL}_{\text{Ref}}$ and GLIO. $\lambda^{\star}_{\texttt{SEC}}$ uses fine-grained labeling [Austin and Flanagan 2009; Rajani and Garg 2018] similar to $\text{GSL}_{\text{Ref}}$, and yet it resembles GLIO in that all runtime labels that come from the syntax are concrete.

### 4.1 Syntax of $\lambda^{\star}_{\texttt{SEC}}$

Our syntax and operations for types are adapted from those of $\text{GSL}_{\text{Ref}}$ and GLIO. Figure 1 defines security labels and security types. For simplicity, we consider base types ({Unit, Bool}), function types, and reference types as our raw types. The PC label decoration $gc$ on a function type

$$
\begin{array}{rcll}
\text{concrete security labels} & \ell, pc & \in & \{\texttt{low}, \texttt{high}\} \\
\text{gradual security labels} & g, gc & ::= & \star \mid \ell \\
\text{base types} & \iota & ::= & \texttt{Unit} \mid \texttt{Bool} \\
\text{raw types} & T, S & ::= & \iota \mid A \xrightarrow{gc} B \mid \texttt{Ref}\ A \\
\text{types} & A, B & ::= & T_g \\
\text{blame labels} & p, q & & \\
\text{variables} & x, y, z & & \\
\text{constants} & k & \in & \{\texttt{unit}, \texttt{true}, \texttt{false}\} \\
\text{terms} & L, M, N & ::= & x \mid (\$\ k)_\ell \mid (\lambda^{pc} x{:}A.\ N)_\ell \mid (L\ M)^p \\
& & \mid & (\texttt{if}\ L\ \texttt{then}\ M\ \texttt{else}\ N)^p \mid \texttt{let}\ x = M\ \texttt{in}\ N \\
& & \mid & (\texttt{ref}\ \ell\ M)^p \mid\ !\ M \mid (L := M)^p \mid (M : A)^p
\end{array}
$$

Fig. 1. Syntax of the surface language $\lambda^\star_{\mathsf{SEC}}$

comes from $\lambda$-abstraction. It is gradual because we allow casting between function types. A raw type forms a type by adding a gradual label ascription. A gradual label can be either concrete ({low, high}) or statically unknown ($\star$).

Figure 1 also defines the syntax of $\lambda^\star_{\mathsf{SEC}}$, with the following characteristics and design choices:

**Concrete runtime labels.** We require concrete security labels (not $\star$) on the syntax of constants $(\$\ k)_\ell$, $\lambda$-abstractions $(\lambda^{pc} x{:}A.\ N)_\ell$, and reference cell creation $(\texttt{ref}\ \ell\ M)^p$. These labels are the mechanism by which the programmer conveys to $\lambda^\star_{\mathsf{SEC}}$ which pieces of data are sensitive and which ones are not. In this way our design choice is similar to GLIO, in which the `toLab` and `new` operators require concrete labels. In contrast, $\mathrm{GSL}_{\mathsf{Ref}}$ allows the programmer to label a value with $\star$, meaning "either low or high", and this design choice is part of why $\mathrm{GSL}_{\mathsf{Ref}}$ violates the DGG. Indeed, we have shown in Section 2.2 a counterexample for the DGG in $\mathrm{GSL}_{\mathsf{Ref}}$, but not in either GLIO or $\lambda^\star_{\mathsf{SEC}}$. To reduce the annotation burden on programmers, we adopt the convention for $\lambda^\star_{\mathsf{SEC}}$ that an unannotated value is shorthand for annotating the value with `low`.

**Type annotations.** There are two places where the programmer may introduce type annotations: 1) the $\lambda$-abstraction $(\lambda^{pc} x{:}A.\ N)_\ell$ and 2) the explicit annotation term $(M : A)^p$. The syntax of types is defined in Figure 1. The programmer has the freedom to control the precision of these annotations and move to either more static or more dynamic, as we have shown in Section 2.

**Support for blame.** To support blame tracking, $\lambda^\star_{\mathsf{SEC}}$ terms that involve implicit casts are decorated with blame labels (in orange) so they can be placed on casts during compilation to $\lambda^{\Rightarrow}_{\mathsf{SEC}}$.

**Labeling granularity.** We choose fine-grained labeling similar to $\mathrm{GSL}_{\mathsf{Ref}}$ because 1) fine-grained labeling and coarse-grained labeling are proved to be equally expressive [Rajani and Garg 2018] 2) fine-grained labeling simplifies the presentation by labeling every value and every type in a uniform way, which declutters the language and makes it easier to study the meta-theory.

**Agda implementation note.** In the Agda development of $\lambda^\star_{\mathsf{SEC}}$, we model terms using abstract binding trees, leveraging the ABT library [1]. We use variable names in this paper for presentation purposes only. In the actual implementation, we employ De Bruijn indices to represent variables and use the ABT library to handle substitution.

### 4.2 Type System of $\lambda^\star_{\mathsf{SEC}}$

The typing rules of $\lambda^\star_{\mathsf{SEC}}$ are shown in Figure 2. The rules are syntax-directed; they are based on the type system of $\mathrm{GSL}_{\mathsf{Ref}}$, which is derived from its static counterpart $\mathrm{SSL}_{\mathsf{Ref}}$ by replacing labels and

[1] https://github.com/jsiek/abstract-binding-trees

$$\boxed{\Gamma; gc \vdash M : A}$$

$$\vdash var\ \frac{\Gamma \ni x : A}{\Gamma; gc \vdash x : A} \qquad \vdash unit\ \frac{}{\Gamma; gc \vdash (\$\ \mathtt{unit})_\ell : \mathtt{Unit}_\ell} \qquad \vdash bool\ \frac{b \in \{\mathtt{true}, \mathtt{false}\}}{\Gamma; gc \vdash (\$\ b)_\ell : \mathtt{Bool}_\ell}$$

$$\vdash lam\ \frac{(\Gamma, x{:}A); pc \vdash N : B}{\Gamma; gc \vdash (\lambda^{pc} x{:}A.\ N)_\ell : (A \xrightarrow{pc} B)_\ell} \qquad \vdash app\ \frac{\begin{array}{ccc}\Gamma; gc \vdash L : (A \xrightarrow{gc'} B)_g & & \Gamma; gc \vdash M : A' \\ A' \lesssim A & g \lesssim gc' & gc \lesssim gc'\end{array}}{\Gamma; gc \vdash (L\ M)^p : B\ \widetilde{\vee}\ g}$$

$$\vdash let\ \frac{\begin{array}{c}\Gamma; gc \vdash M : A \\ (\Gamma, x{:}A); gc \vdash N : B\end{array}}{\Gamma; gc \vdash \mathtt{let}\ x = M\ \mathtt{in}\ N : B} \qquad \vdash if\ \frac{\begin{array}{c}\Gamma; gc \vdash L : \mathtt{Bool}_g \\ \Gamma; gc\ \widetilde{\vee}\ g \vdash M : A \qquad \Gamma; gc\ \widetilde{\vee}\ g \vdash N : B \\ A\ \widetilde{\vee}\ B = C\end{array}}{\Gamma; gc \vdash (\mathtt{if}\ L\ \mathtt{then}\ M\ \mathtt{else}\ N)^p : C\ \widetilde{\vee}\ g}$$

$$\vdash ref\ \frac{\begin{array}{c}\Gamma; gc \vdash M : T_g \\ T_g \lesssim T_\ell \qquad gc \lesssim \ell\end{array}}{\Gamma; gc \vdash (\mathtt{ref}\ \ell\ M)^p : (\mathtt{Ref}\ T_\ell)_{\mathtt{low}}} \qquad \vdash deref\ \frac{\Gamma; gc \vdash M : (\mathtt{Ref}\ A)_g}{\Gamma; gc \vdash !\ M : A\ \widetilde{\vee}\ g}$$

$$\vdash assign\ \frac{\begin{array}{ccc}\Gamma; gc \vdash L : (\mathtt{Ref}\ T_{\hat{g}})_g & & \Gamma; gc \vdash M : A \\ A \lesssim T_{\hat{g}} & g \lesssim \hat{g} & gc \lesssim \hat{g}\end{array}}{\Gamma; gc \vdash (L := M)^p : \mathtt{Unit}_{\mathtt{low}}} \qquad \vdash ann\ \frac{\begin{array}{c}\Gamma; gc \vdash M : A' \\ A' \lesssim A\end{array}}{\Gamma; gc \vdash (M : A)^p : A}$$

Fig. 2. Typing rules of the surface language $\lambda^\star_{\mathsf{SEC}}$

types as well as their operators and predicates with the gradual variants. $\mathrm{SSL}_{\mathsf{Ref}}$ is in turn a straightforward adaptation of prior security-typed languages (Fennell and Thiemann [2013]; Heintze and Riecke [1998]; Zdancewic [2002]).

For example, in $\mathrm{SSL}_{\mathsf{Ref}}$ the typing rule of application looks like:

$$\frac{\begin{array}{ccc}\Gamma; pc \vdash L : (A \xrightarrow{pc'} B)_\ell & & \Gamma; pc \vdash M : A' \\ A' <: A & \ell \preccurlyeq pc' & pc \preccurlyeq pc'\end{array}}{\Gamma; pc \vdash (L\ M) : B \vee \ell} \tag{7}$$

where $A' <: A$ is the usual type subsumption of function argument. The side conditions $\ell \preccurlyeq pc'$ and $pc \preccurlyeq pc'$ restricts the PC label on the function type so that no information is leaked through side effects. The type of the application has label that is the join of the label on $B$ and $\ell$ ($B \vee \ell$). In $\lambda^\star_{\mathsf{SEC}}$, the typing judgment takes the form $\Gamma; gc \vdash M : A$, where the static PC $gc$ and the type $A$ become gradual (may be or contain $\star$). Like $\mathrm{GSL}_{\mathsf{Ref}}$, we replace label partial order with label consistent subtyping, type subtyping with type consistent subtyping, and label join with label consistent join and get rule $\vdash app$. Similarly, in $\vdash if$ the join of the types from the two branches is replace by the consistent join $A\ \widetilde{\vee}\ B$. We define the gradual predicates and operators in Figure 9 and Figure 10 in the Appendix.[2] They are straightforwardly adapted from those of $\mathrm{GSL}_{\mathsf{Ref}}$ and GLIO.

The only major difference from the type system of $\mathrm{GSL}_{\mathsf{Ref}}$ is that because of the concrete label restriction on the syntax of constants and $\lambda$-abstractions, these terms must have concrete labels at the top level of their respective types (rule $\vdash unit$, $\vdash bool$, and $\vdash lam$). Similarly, the type of the value in a newly allocated cell (rule $\vdash$*ref*) has a concrete top-level label: $(\texttt{Ref}\ T_\ell)_{\texttt{low}}$. The reference itself has a low label because it is newly created and cannot leak information.

[2]Note to reviewers: the Appendix of this paper is in the supplementary text.

| | | | |
|---|---|---|---|
| errors | $e$ | ::= | $\texttt{nsu-error} \mid \texttt{blame}^{p}$ |
| casts | $c$ | ::= | $A \Rightarrow^{p} B$ |
| terms | $L, M, N$ | ::= | $x \mid (\$\, k)_\ell \mid (\texttt{addr}\ a)_\ell \mid (\lambda^{pc} x{:}A.\, N)_\ell \mid L\ M \mid \texttt{if}\ L\ A\ M\ N$ |
| | | \| | $\texttt{let}\ x = M\ \texttt{in}\ N \mid \texttt{ref}\ \ell\ M \mid \texttt{ref}^{\checkmark}\ \ell\ M \mid \texttt{ref}^{?}\ \ell\ M \mid\ !\ M$ |
| | | \| | $L := M \mid L :=^{\checkmark} M \mid L :=^{?} M \mid M \langle c \rangle \mid \texttt{cast}_{\texttt{pc}}\ g\ M$ |
| | | \| | $\texttt{prot}\ \ell\ M \mid \texttt{error}\ e \mid \bullet$ |
| values | $V$ | ::= | $(\texttt{addr}\ a)_\ell \mid (\$\, k)_\ell \mid (\lambda^{pc} x{:}A.\, N)_\ell \mid \bullet \mid V \langle c \rangle$ , where $c$ is inert |

Fig. 3. Syntax of the cast calculus $\lambda^{\Rightarrow}_{\mathsf{SEC}}$

## 5 $\lambda^{\Rightarrow}_{\mathsf{SEC}}$: THE CAST CALCULUS (CC)

In this section we present the cast calculus $\lambda^{\Rightarrow}_{\mathsf{SEC}}$. We define the syntax (Section 5.1), the type system (Section 5.3), and the operational semantics for $\lambda^{\Rightarrow}_{\mathsf{SEC}}$ (Section 5.4). We show that $\lambda^{\star}_{\mathsf{SEC}}$ can be compiled into $\lambda^{\Rightarrow}_{\mathsf{SEC}}$ by inserting casts and NSU checks in a type-directed way (Section 5.2).

### 5.1 Syntax of $\lambda^{\Rightarrow}_{\mathsf{SEC}}$

The syntax of $\lambda^{\Rightarrow}_{\mathsf{SEC}}$ is shown in Figure 3. Compared with the surface language $\lambda^{\star}_{\mathsf{SEC}}$, $\lambda^{\Rightarrow}_{\mathsf{SEC}}$ has the following auxiliary language constructs:

**Explicit casts.** Casts are made explicit using the cast term $M \langle c \rangle$. A cast $c$ is of shape $A \Rightarrow^{p} B$, where $A$ is the source type, $B$ the target type, and $p$ is the blame label. We require that $A$ is *consistent* with $B$, written $A \sim B$, defined in Figure 8 of the Appendix.

**Support for NSU checking.** Recall that NSU checking is required to prevent illegal implicit flows through the heap whenever static typing information is insufficient to decide whether a heap write operation is safe or not. Consequently, in $\lambda^{\Rightarrow}_{\mathsf{SEC}}$, we have variants of the reference creation and assignment terms that statically prevent illegal implicit flows, $\texttt{ref}\ \ell\ M$ and $L := M$, and other variants that dynamically prevent illegal implicit flows using NSU, $\texttt{ref}^{?}\ \ell\ M$ and $L :=^{?} M$. During reduction, a statically-enforced heap write operation immediately becomes a checked write, $\texttt{ref}^{\checkmark}\ M$ or $L :=^{\checkmark} M$. On the other hand, the dynamic (NSU) variant reduces to the checked form or throws an NSU error depending on whether the NSU check passes.

**Terms that arise during reduction.** As usual, we have an address term $(\texttt{addr}\ a)_\ell$. It has a label decoration $\ell$ just like constants and $\lambda$, which indicates the security level of the address itself. As we have discussed, an address is of shape $a = n_{\hat{\ell}}$, where $\hat{\ell}$ signifies which half of the heap, low or high, the address points to. In addition, we have an error term $\texttt{error}\ e$, where $e$ can be either 1) a cast failure $\texttt{blame}^{p}$ or 2) an nsu-error due to a failed NSU check. The protection term $\texttt{prot}\ \ell\ M$ ensures that the computation result of $M$ and the side effects in $M$ must be at least as secure as $\ell$. Finally, the term $\texttt{cast}_{\texttt{pc}}\ g\ M$ (PC cast) is useful for ensuring type preservation. It can be viewed as an adapter between the inner and the outer static PCs, as it uses the security label $g$ as the static PC to type check the sub-term $M$. The $\texttt{cast}_{\texttt{pc}}\ g$ goes away as soon as $M$ reduces to a value.

**"Opaque" term for noninterference proof.** We use $\bullet$ for the opaque, erased value in $\lambda^{\Rightarrow}_{\mathsf{SEC}}$.

The values of $\lambda^{\Rightarrow}_{\mathsf{SEC}}$ include addresses, constants, functions, opaque values, and values wrapped in an inert cast. To explain the latter, we categorize casts into *active* casts, which can be applied and then further reduce, and *inert* casts, which are value-forming (Figure 13 of the Appendix). If a value contains at least one inert cast, we say that it is *wrapped*. We refer to wrapped functions and references as *function proxies* and *reference proxies* respectively.

### 5.2 Compilation from $\lambda^{\star}_{\mathsf{SEC}}$ to $\lambda^{\Rightarrow}_{\mathsf{SEC}}$

The function $C$ compiles a well-typed $\lambda^{\star}_{\mathsf{SEC}}$ programs into $\lambda^{\Rightarrow}_{\mathsf{SEC}}$; it is shown in full in Figure 12 of the Appendix. The main idea is that we insert casts whenever there are consistent subtyping side conditions on a $\lambda^{\star}_{\mathsf{SEC}}$ typing rule. To obtain the source and target types of the casts, which must satisfy *consistency* instead of consistent subtyping, we turn to the *merge operators*, written $\leftharpoondown$, between labels or types [Siek and Taha 2007]. The merge operators decouple consistency from subtyping: the $\leftharpoondown$ operator takes two types $A, B$ that satisfy $A \lesssim B$ and calculates $C$ such that $A \sim C <: B$, while the $\leftharpoondown$ operator, defined dually, calculates $C'$ such that $A <: C' \sim B$. The definitions of the merge operators are shown in Figure 11 of the Appendix and the subtyping relations are defined in Figure 7 of the Appendix. Now consider the typing rule $\vdash$*app* which requires the argument type $A'$ to be a consistent subtype of the function's parameter type $A$, so we insert a cast on the argument from $A'$ to $A' \leftharpoondown A$. The $\vdash$*if* case requires extra attention, because it contains consistent join. It can be converted into two uses of consistent subtyping because $A \mathbin{\widetilde{\vee}} B = C$ implies $A \lesssim C$ and $B \lesssim C$. We insert casts from $A$ to $A \leftharpoondown C$ and from $B$ to $B \leftharpoondown C$ in the two branches.

There are two consistent subtyping side conditions that are different from the rest, on reference creation ($\vdash$*ref*) and assignment ($\vdash$*assign*), which we highlight in Figure 2. They are *not* compiled into casts. Instead, they decide whether we perform NSU checking or not. If both labels in each side condition are concrete, we skip NSU checking by compiling into the statically-enforced variant, `ref` $\ell$ $M$ or $L$ `:=` $M$; otherwise, we perform runtime NSU checking by compiling into `ref`$^{?}$ $\ell$ $M$ or $L$ `:=`$^{?}$ $M$. In this way, unlike $\mathrm{GSL}_{\mathsf{Ref}}$ or GLIO, the runtime overhead of NSU checks is only incurred in dynamically-typed regions of code, and not statically-typed regions.

### 5.3 Type System of $\lambda^{\Rightarrow}_{\mathsf{SEC}}$

Figure 4 shows the typing rules for $\lambda^{\Rightarrow}_{\mathsf{SEC}}$. The typing judgment is of form $\Gamma; \Sigma; gc; pc \vdash M : A$. $\Gamma$ and $gc$ have the same meanings as in the typing of $\lambda^{\star}_{\mathsf{SEC}}$. $\Sigma$ is the heap typing context and $pc$ is the dynamic PC, both of which play an import role during reduction.

The heap context is split into low- and high- halves just like the heap:

$$\Sigma = \langle \Sigma_{\mathtt{low}}, \Sigma_{\mathtt{high}} \rangle \text{ , where } \Sigma_{\mathtt{low}}, \Sigma_{\mathtt{high}} : \mathit{List}\ (\mathit{Index} \times \mathit{RawType})$$

where each half is an association list from indices, modeled by $\mathbb{N}$, to raw types. The type $A$ that corresponds to a certain address $a = n_{\ell}$ is looked-up by $\Sigma(n_{\ell}) = (\Sigma_{\ell}(n))_{\ell}$ where $\Sigma_{\ell}(n)$ is usual association list indexing. The type that an address references remains unchanged during reduction, so reference creation is the only occasion that $\Sigma$ grows. We write $\emptyset = \langle [], [] \rangle$ as a shorthand (for both halves being empty).

A novel feature of $\lambda^{\Rightarrow}_{\mathsf{SEC}}$'s type system is that we quantify the typing judgment by $pc$ to capture successful NSU checks. In the typing rules for checked reference creation ($\vdash$*ref*✓) and assignment ($\vdash$*assign*✓), there are side conditions $pc \preccurlyeq \ell$, highlighted in Figure 4, capturing the heap policy that the dynamic PC $pc$ is a lower bound on the security of all memory locations that are written to, which is enforced by the NSU checks at runtime. Another novel aspect of rules is that in the premises for sub-terms that do not immediately reduce, such as the body of a $\lambda$ and the branches of an if-expression, we universally quantify the $pc$ (as in $\forall pc$). This universal quantification helps us prove "compilation preserves type" (Lemma 9), because the typing judgment of $\lambda^{\star}_{\mathsf{SEC}}$ does not contain $pc$ while the typing judgment of $\lambda^{\Rightarrow}_{\mathsf{SEC}}$ does. The reason this quantification being okay is that we only insert static and NSU variants of heap write operations during compilation, while the

$$\boxed{\Gamma; \Sigma; gc; pc \vdash M : A}$$

$$\vdash var \; \frac{\Gamma \ni x : A}{\Gamma; \Sigma; gc; pc \vdash x : A} \qquad \vdash unit \; \frac{}{\Gamma; \Sigma; gc; pc \vdash (\$ \; \mathtt{unit})_\ell : \mathtt{Unit}_\ell}$$

$$\vdash bool \; \frac{b \in \{\mathtt{true}, \mathtt{false}\}}{\Gamma; \Sigma; gc; pc \vdash (\$ \; b)_\ell : \mathtt{Bool}_\ell} \qquad \vdash addr \; \frac{\Sigma(a) = A}{\Gamma; \Sigma; gc; pc \vdash (\mathtt{addr} \; a)_\ell : (\mathtt{Ref} \; A)_\ell}$$

$$\vdash lam \; \frac{\forall pc''. (\Gamma, x{:}A); \Sigma; pc'; pc'' \vdash N : B}{\Gamma; \Sigma; gc; pc \vdash (\lambda^{pc'} x{:}A.\, N)_\ell : (A \xrightarrow{pc'} B)_\ell} \qquad \vdash let \; \frac{\Gamma; \Sigma; gc; pc \vdash M : A \quad \forall pc'. (\Gamma, x{:}A); \Sigma; gc; pc' \vdash N : B}{\Gamma; \Sigma; gc; pc \vdash \mathtt{let} \; x = M \; \mathtt{in} \; N : B}$$

$$\vdash app \; \frac{\Gamma; \Sigma; gc; pc \vdash L : (A \xrightarrow{gc \,\tilde{\vee}\, g} B)_g \quad \Gamma; \Sigma; gc; pc \vdash M : A}{\Gamma; \Sigma; gc; pc \vdash L \; M : B \,\tilde{\vee}\, g} \qquad \vdash if \; \frac{\Gamma; \Sigma; gc; pc \vdash L : \mathtt{Bool}_g \quad \forall pc'. \Gamma; \Sigma; gc \,\tilde{\vee}\, g; pc' \vdash M : A \quad \forall pc'. \Gamma; \Sigma; gc \,\tilde{\vee}\, g; pc' \vdash N : A}{\Gamma; \Sigma; gc; pc \vdash \mathtt{if} \; L \; A \; M \; N : A \,\tilde{\vee}\, g}$$

$$\vdash ref \; \frac{\Gamma; \Sigma; pc'; pc \vdash M : T_\ell \quad pc' \preccurlyeq \ell}{\Gamma; \Sigma; pc'; pc \vdash \mathtt{ref} \; \ell \; M : (\mathtt{Ref} \; T_\ell)_{\mathtt{low}}} \qquad \vdash ref\checkmark \; \frac{\Gamma; \Sigma; gc; pc \vdash M : T_\ell \quad pc \preccurlyeq \ell}{\Gamma; \Sigma; gc; pc \vdash \mathtt{ref}^{\checkmark} \; \ell \; M : (\mathtt{Ref} \; T_\ell)_{\mathtt{low}}}$$

$$\vdash ref? \; \frac{\Gamma; \Sigma; gc; pc \vdash M : T_\ell}{\Gamma; \Sigma; gc; pc \vdash \mathtt{ref}^{?} \; \ell \; M : (\mathtt{Ref} \; T_\ell)_{\mathtt{low}}} \qquad \vdash deref \; \frac{\Gamma; \Sigma; gc; pc \vdash M : (\mathtt{Ref} \; A)_g}{\Gamma; \Sigma; gc; pc \vdash \, ! \; M : A \,\tilde{\vee}\, g}$$

$$\vdash assign \; \frac{\Gamma; \Sigma; pc'; pc \vdash L : (\mathtt{Ref} \; T_\ell)_\ell \quad \Gamma; \Sigma; pc'; pc \vdash M : T_\ell \quad pc' \preccurlyeq \ell}{\Gamma; \Sigma; pc'; pc \vdash L \; \mathtt{:=} \; M : \mathtt{Unit}_{\mathtt{low}}} \qquad \vdash assign\checkmark \; \frac{\Gamma; \Sigma; gc; pc \vdash L : (\mathtt{Ref} \; T_\ell)_\ell \quad \Gamma; \Sigma; gc; pc \vdash M : T_\ell \quad pc \preccurlyeq \ell}{\Gamma; \Sigma; gc; pc \vdash L \; \mathtt{:=}^{\checkmark} \; M : \mathtt{Unit}_{\mathtt{low}}}$$

$$\vdash assign? \; \frac{\Gamma; \Sigma; gc; pc \vdash L : (\mathtt{Ref} \; T_g)_g \quad \forall pc'. \Gamma; \Sigma; gc; pc' \vdash M : T_g}{\Gamma; \Sigma; gc; pc \vdash L \; \mathtt{:=}^{?} \; M : \mathtt{Unit}_{\mathtt{low}}}$$

$$\vdash cast \; \frac{\Gamma; \Sigma; gc; pc \vdash M : A}{\Gamma; \Sigma; gc; pc \vdash M \langle A \Rightarrow^{p} B \rangle : B} \qquad \vdash cast_{pc} \; \frac{\Gamma; \Sigma; g; pc \vdash M : A \quad pc \sim g}{\Gamma; \Sigma; gc; pc \vdash \mathtt{cast}_{\mathtt{pc}} \; g \; M : A}$$

$$\vdash prot \; \frac{\Gamma; \Sigma; gc \,\tilde{\vee}\, \ell; pc \vee \ell \vdash M : A}{\Gamma; \Sigma; gc; pc \vdash \mathtt{prot} \; \ell \; M : A \,\tilde{\vee}\, \ell} \qquad \vdash error \; \frac{}{\Gamma; \Sigma; gc; pc \vdash \mathtt{error} \; e : A}$$

$$\vdash sub \; \frac{\Gamma; \Sigma; gc; pc \vdash M : A \quad A <: B}{\Gamma; \Sigma; gc; pc \vdash M : B} \qquad \vdash sub_{pc} \; \frac{\Gamma; \Sigma; gc'; pc \vdash M : A \quad gc <: gc'}{\Gamma; \Sigma; gc; pc \vdash M : A}$$

Fig. 4. Typing rules of the cast calculus $\lambda_{\mathtt{SEC}}^{\Rightarrow}$

checked variants do not appear until reduction. If one sub-term has not yet been reduced, there is no checked heap writes in it, so there is no term setting any constraint on *pc*.

Rule $\vdash$*ref* and rule $\vdash$*assign* perform static enforcement of the heap policy, highlighted in Figure 4. In these two rules, we use $pc' \preccurlyeq \ell$ as side condition, where $gc = pc'$ is a concrete static PC. This is because during compilation we only insert the static variants of heap writes when $gc$ is concrete. We are going to show in Section 5.4 that this static check supersedes its dynamic counterpart, i.e., NSU checking. The NSU rules $\vdash$*ref*? and $\vdash$*assign*? do not have any side condition. They are for situations in which the heap policy *will be* dynamically enforced by NSU, but the actual check has not yet happened.

Rule $\vdash$*cast* captures type consistency. The cast $A \Rightarrow^{p} B$ casts term $M$ from type $A$ to $B$, where $A \sim B$. Rule $\vdash cast_{pc}$ allows us to switch from the current static PC $gc$, to a label $g$ consistent with the current dynamic PC to type the sub-term. It is useful for proving type safety. Rule $\vdash$*sub* is the subsumption rule for types. When we discharge consistent subtyping into consistency and subtyping, all subtyping relations are collapsed into this single rule. Similarly $\vdash sub_{pc}$ is the subsumption rule for static PCs. The intuition is that if $gc'$ is a lower bound of all side effects in $M$, then $gc$, which is lower, must also be a lower bound.

Rule $\vdash$*prot* is in accordance to the semantics of the protection term: the computation result is protected at $\ell$, thus the type is stamped with $\ell$; all side effects in its sub-term $M$ must write to memory locations at least as secure as $\ell$, so both PCs typing $M$ are also stamped with $\ell$.

### 5.4 Small-step and Big-step Operational Semantics

In this section, we first present a small-step semantics for $\lambda^{\Rightarrow}_{\mathsf{SEC}}$, which defines the dynamic semantics for $\lambda^{\Rightarrow}_{\mathsf{SEC}}$. After that, we define a big-step semantics that we use as a technical device to prove noninterference.

The small-step relation is of form $M \mid \mu \mid pc \longrightarrow M' \mid \mu'$, which reduces the configuration of term $M$, heap $\mu$ under dynamic PC $pc$ to another configuration $M'$, $\mu'$. We formally define heap $\mu$ as a pair of association lists mapping indices to values, one for low and the other for high:

$$\mu = \langle \mu_{\mathtt{low}}, \mu_{\mathtt{high}} \rangle \text{ , where } \mu_{\mathtt{low}}, \mu_{\mathtt{high}} : List\ (Index \times Value)$$

The lookup and cons of $\mu$ are defined using the respective operations for association lists after performing case analysis on the address's label:

$$lookup\ \mu\ n_{\ell} = \mu_{\ell}(n) \qquad \text{and} \qquad cons\ n_{\ell}\ V\ \mu = \begin{cases} \langle \langle n, V \rangle :: \mu_{\mathtt{low}}, \mu_{\mathtt{high}} \rangle \text{ , if } \ell = \mathtt{low} \\ \langle \mu_{\mathtt{low}}, \langle n, V \rangle :: \mu_{\mathtt{high}} \rangle \text{ , if } \ell = \mathtt{high} \end{cases}$$

An address $a = n_{\ell}$ is fresh if and only if its index $n$ equals to the length of the association list that represents the half-heap of $\ell$. Whenever we create a new reference or perform an assignment, the heap grows by one index-value pair. There can be multiple pairs that contain the same index, while the definition of *lookup* ensures that we always get back the latest value that an address references. The shorthand for the empty heap is $\emptyset = \langle [], [] \rangle$.

We represent evaluation contexts using frames (Figure 16 of the Appendix). The *plug* function replaces the hole ($\square$) in a frame with a term and produces a term. In this way, all congruence reduction rules are collapsed into a single $\xi$ rule (Figure 5) using *plug*. Similarly, $\xi$-*err* propagates an error $e$ outside a frame.

Let us get back to Figure 5. Conforming to the usual approach of IFC, the protection term has two functionalities: 1) it protects the computation result by stamping $\ell$ on the result value (rule *prot-val*) 2) it limits the side effects in its sub-term $M$ to be at least as secure as $\ell$, by upgrading the PC used to reduce $M$ to $pc \vee \ell$ (rule *prot-ctx*). Also standard, we insert protection terms in

$$\boxed{M \mid \mu \mid pc \longrightarrow M' \mid \mu'}$$

$$\xi \; \frac{M \mid \mu \mid pc \longrightarrow M' \mid \mu'}{plug\ M\ F \mid \mu \mid pc \longrightarrow plug\ M'\ F \mid \mu'} \qquad \xi\text{-}err \; \frac{}{plug\ (\mathtt{error}\ e)\ F \mid \mu \mid pc \longrightarrow \mathtt{error}\ e \mid \mu}$$

$$prot\text{-}val \; \frac{}{\mathtt{prot}\ \ell\ V \mid \mu \mid pc \longrightarrow V \curlyvee \ell \mid \mu} \qquad prot\text{-}ctx \; \frac{M \mid \mu \mid pc \curlyvee \ell \longrightarrow M' \mid \mu'}{\mathtt{prot}\ \ell\ M \mid \mu \mid pc \longrightarrow \mathtt{prot}\ \ell\ M' \mid \mu'}$$

$$prot\text{-}err \; \frac{}{\mathtt{prot}\ \ell\ (\mathtt{error}\ e) \mid \mu \mid pc \longrightarrow \mathtt{error}\ e \mid \mu}$$

$$\beta \; \frac{}{(\lambda^{pc'} x{:}A.\ N)_\ell\ V \mid \mu \mid pc \longrightarrow \mathtt{prot}\ \ell\ (N[x := V]) \mid \mu}$$

$$\beta\text{-}if\text{-}true \; \frac{}{\mathtt{if}\ (\$\ \mathtt{true})_\ell\ A\ M\ N \mid \mu \mid pc \longrightarrow \mathtt{prot}\ \ell\ M \mid \mu}$$

$$\beta\text{-}if\text{-}false \; \frac{}{\mathtt{if}\ (\$\ \mathtt{false})_\ell\ A\ M\ N \mid \mu \mid pc \longrightarrow \mathtt{prot}\ \ell\ N \mid \mu}$$

$$\beta\text{-}let \; \frac{}{\mathtt{let}\ x = V\ \mathtt{in}\ N \mid \mu \mid pc \longrightarrow N[x := V] \mid \mu} \qquad ref\text{-}static \; \frac{}{\mathtt{ref}\ \ell\ M \mid \mu \mid pc \longrightarrow \mathtt{ref}^{\checkmark}\ \ell\ M \mid \mu}$$

$$ref?\text{-}ok \; \frac{pc \preccurlyeq \ell}{\mathtt{ref}^{?}\ \ell\ M \mid \mu \mid pc \longrightarrow \mathtt{ref}^{\checkmark}\ \ell\ M \mid \mu} \qquad ref?\text{-}fail \; \frac{pc \not\preccurlyeq \ell}{\mathtt{ref}^{?}\ \ell\ M \mid \mu \mid pc \longrightarrow \mathtt{error}\ \mathtt{nsu{-}error} \mid \mu}$$

$$ref \; \frac{a = n_\ell\ \textbf{FreshIn}\ \mu}{\mathtt{ref}^{\checkmark}\ \ell\ V \mid \mu \mid pc \longrightarrow (\mathtt{addr}\ a)_{\mathtt{low}} \mid cons\ a\ V\ \mu}$$

$$deref \; \frac{lookup\ \mu\ a = V}{!\ (\mathtt{addr}\ a)_\ell \mid \mu \mid pc \longrightarrow \mathtt{prot}\ (\hat{\ell} \curlyvee \ell)\ V \mid \mu}\text{, where } a = n_{\hat{\ell}}$$

$$assign\text{-}static \; \frac{}{L := M \mid \mu \mid pc \longrightarrow L :=^{\checkmark} M \mid \mu}$$

$$assign?\text{-}ok \; \frac{pc \preccurlyeq \hat{\ell}}{(\mathtt{addr}\ a)_\ell :=^{?} M \mid \mu \mid pc \longrightarrow (\mathtt{addr}\ a)_\ell :=^{\checkmark} M \mid \mu}\text{, where } a = n_{\hat{\ell}}$$

$$assign?\text{-}fail \; \frac{pc \not\preccurlyeq \hat{\ell}}{(\mathtt{addr}\ a)_\ell :=^{?} M \mid \mu \mid pc \longrightarrow \mathtt{error}\ \mathtt{nsu{-}error} \mid \mu}\text{, where } a = n_{\hat{\ell}}$$

$$assign \; \frac{}{(\mathtt{addr}\ a)_\ell :=^{\checkmark} V \mid \mu \mid pc \longrightarrow (\$\ \mathtt{unit})_{\mathtt{low}} \mid cons\ a\ V\ \mu}$$

Fig. 5. Small-step operational semantics for $\lambda_{\mathsf{SEC}}^{\Rightarrow}$

*β-if* and *β*, thus preventing implicit flows from the branch condition, or from which function is being applied. We next introduce the more interesting reduction rules of $\lambda_{\mathsf{SEC}}^{\Rightarrow}$, which fall into two categories: 1) the ones about the heap 2) the ones that deal with casts.

The rules for heap operations can be divided into reading and writing. Reading (rule *deref*) is simple: we protect the value looked-up from the heap with two labels: 1) $\ell$ on the address term, to prevent leaking which address is being dereferenced (analogous to $\beta$) 2) $\hat{\ell}$, to ensure that the value

$$\boxed{M \mid \mu \mid pc \longrightarrow M' \mid \mu'}$$

$$\beta\text{-}cast\text{-}pc\ \frac{}{\texttt{cast}_{\texttt{pc}}\ g\ V \mid \mu \mid pc \longrightarrow V \mid \mu} \qquad cast\ \frac{\textbf{Active}\ c \quad \textbf{Cast}\ V\ ,\ c \rightsquigarrow M}{V\ \langle\, c\,\rangle \mid \mu \mid pc \longrightarrow M \mid \mu}$$

$$if\text{-}cast\text{-}true\ \frac{\textbf{Inert}\ c}{\texttt{if}\ ((\$\ \texttt{true})_\ell\ \langle\, c\,\rangle)\ A\ M\ N \mid \mu \mid pc \longrightarrow (\texttt{prot}\ \ell\ (\texttt{cast}_{\texttt{pc}}\ \star\ M))\ \langle\, branch_c\ A\ c\,\rangle \mid \mu}$$

$$if\text{-}cast\text{-}false\ \frac{\textbf{Inert}\ c}{\texttt{if}\ ((\$\ \texttt{false})_\ell\ \langle\, c\,\rangle)\ A\ M\ N \mid \mu \mid pc \longrightarrow (\texttt{prot}\ \ell\ (\texttt{cast}_{\texttt{pc}}\ \star\ N))\ \langle\, branch_c\ A\ c\,\rangle \mid \mu}$$

$$fun\text{-}cast\ \frac{\textbf{Inert}\ c}{(V\ \langle\, c\,\rangle)\ W \mid \mu \mid pc \longrightarrow elim\text{-}fun\text{-}proxy\ V\ W\ c\ pc \mid \mu}$$

$$deref\text{-}cast\ \frac{\textbf{Inert}\ c}{!\ (V\ \langle\, c\,\rangle) \mid \mu \mid pc \longrightarrow (!\ V)\ \langle\, out_c\ c\,\rangle \mid \mu}$$

$$assign?\text{-}cast\ \frac{\textbf{Inert}\ c}{(V\ \langle\, c\,\rangle)\ :=^?\ M \mid \mu \mid pc \longrightarrow elim\text{-}ref\text{-}proxy\ V\ M\ c\ \text{-}{:=^?}\text{-} \mid \mu}$$

$$assign\text{-}cast\ \frac{\textbf{Inert}\ c}{(V\ \langle\, c\,\rangle)\ :=^{\checkmark}\ W \mid \mu \mid pc \longrightarrow elim\text{-}ref\text{-}proxy\ V\ W\ c\ \text{-}{:=^{\checkmark}}\text{-} \mid \mu}$$

$$\boxed{\begin{array}{l} branch_c : Type \to Cast \to Cast \\ dom_c, cod_c, in_c, out_c : Cast \to Cast \end{array}}$$

$$branch_c\ A\ (\texttt{Bool}_g \Rightarrow^p \texttt{Bool}_\star) = A \mathbin{\widetilde{\curlyvee}} g \Rightarrow^p A \mathbin{\widetilde{\curlyvee}} \star \tag{8}$$

$$\begin{aligned}
dom_c\ ((A \xrightarrow{gc_1} B)_{g_1} \Rightarrow^p (C \xrightarrow{gc_2} D)_{g_2}) &= C \Rightarrow^p A \\
cod_c\ ((A \xrightarrow{gc_1} B)_{g_1} \Rightarrow^p (C \xrightarrow{gc_2} D)_{g_2}) &= B \mathbin{\widetilde{\curlyvee}} g_1 \Rightarrow^p D \mathbin{\widetilde{\curlyvee}} g_2 \\
in_c\ ((\texttt{Ref}\ A)_{g_1} \Rightarrow^p (\texttt{Ref}\ B)_{g_2}) &= B \Rightarrow^p A \\
out_c\ ((\texttt{Ref}\ A)_{g_1} \Rightarrow^p (\texttt{Ref}\ B)_{g_2}) &= A \mathbin{\widetilde{\curlyvee}} g_1 \Rightarrow^p B \mathbin{\widetilde{\curlyvee}} g_2
\end{aligned}$$

$$\boxed{\begin{array}{l} elim\text{-}fun\text{-}proxy : Term \to Term \to (c : Cast) \to (pc : ConcreteLabel) \to Term \\ elim\text{-}ref\text{-}proxy : Term \to Term \to (c : Cast) \to (\text{-}{:=^\dagger}\text{-} \in \{\text{-}{:=}\text{-}, \text{-}{:=^?}\text{-}, \text{-}{:=^\checkmark}\text{-}\}) \to Term \end{array}}$$

$$elim\text{-}fun\text{-}proxy\ V\ W\ ((A \xrightarrow{pc_1} B)_{\ell_1} \Rightarrow^p (C \xrightarrow{pc_2} D)_{g_2})\ pc = (V\ (W\ \langle\, dom_c\ c\,\rangle))\ \langle\, cod_c\ c\,\rangle \tag{9}$$

$$elim\text{-}fun\text{-}proxy\ V\ W\ ((A \xrightarrow{pc_1} B)_{\ell_1} \Rightarrow^p (C \xrightarrow{\star} D)_{g_2})\ pc = \begin{cases} (\texttt{cast}_{\texttt{pc}}\ pc\ (V\ (W\ \langle\, dom_c\ c\,\rangle)))\ \langle\, cod_c\ c\,\rangle \\ \quad ,\ \text{if}\ pc \curlyvee \ell_1 \preccurlyeq pc_1 \\ \texttt{error}\ \texttt{blame}^p\ ,\ \text{otherwise} \end{cases} \tag{10}$$

$$elim\text{-}ref\text{-}proxy\ V\ M\ ((\texttt{Ref}\ S_{\hat{\ell}_1})_\ell \Rightarrow^p (\texttt{Ref}\ T_{\hat{\ell}_2})_g)\ \text{-}{:=^\dagger}\text{-} = V\ :=^\dagger\ (M\ \langle\, in_c\ c\,\rangle)$$

$$elim\text{-}ref\text{-}proxy\ V\ M\ ((\texttt{Ref}\ S_{\hat{\ell}_1})_\ell \Rightarrow^p (\texttt{Ref}\ T_\star)_g)\ \text{-}{:=^\dagger}\text{-} = \begin{cases} V\ :=^\dagger\ (M\ \langle\, in_c\ c\,\rangle)\ ,\ \text{if}\ \ell \preccurlyeq \hat{\ell}_1 \\ \texttt{error}\ \texttt{blame}^p\ ,\ \text{otherwise} \end{cases}$$

Fig. 6. Small-step operational semantics cont'd: elimination rules for casts

read from the high-heap is always high-security. Writing involves rules of three forms: static, not-yet-checked, and checked. Consider reference creation. Rule *ref-static* goes from the static form `ref` $\ell$ $M$ to the checked form $\texttt{ref}^{\checkmark}$ $\ell$ $M$ directly. Reduction preserves type (Section 6.1); therefore, the side condition $pc' \preccurlyeq \ell$ on $\vdash ref$ (Figure 4) supersedes $pc \preccurlyeq \ell$ on $\vdash ref\checkmark$, so no NSU checking is required. Rule *ref?-ok* performs a successful NSU check and reduces the not-yet-checked ($\texttt{ref}^{?}$ $\ell$ $M$) to checked. When proving type preservation, the check $pc \preccurlyeq \ell$ goes into the typing of the checked term (rule $\vdash ref\checkmark$). If NSU fails, we use *ref?-fail* and go to `nsu-error`. Finally, rule *ref* reduces the checked form, creates a fresh memory location, and returns the address of the new location. The address term has label `low` because the new address is freshly allocated. Assignment follows the same pattern, the caveat being that the label $\hat{\ell}$ used in the NSU checks in rule *assign?-ok* and rule *assign?-fail* comes from the address instead of the term.

The reduction rules that involve casts are shown in Figure 6. Applying an active cast is summarized in a single rule *cast* utilizing relation **ApplyCast** (Figure 14 of the Appendix). We briefly describe **ApplyCast**. Identity casts of base types are discarded immediately (*cast-base-id*). When projecting to a base type with $\ell_2$, the value must be of canonical form that contains an injection from $\ell_1$. We check whether $\ell_1$ subsumes $\ell_2$, because of subtyping. If $\ell_1 \preccurlyeq \ell_2$, we discard the injection and the projection at the same time (*cast-base-proj*); if not, we blame the projection (*cast-base-proj-blame*). A function cast is active, if either 1) the label $g_1$ or 2) the PC label $gc_1$ of the source type is $\star$ (*A-fun*, *A-fun-pc*). On the other hand, a function cast is inert if both $g_1$ and $gc_1$ are concrete (*I-fun*). In case (1) the cast is applied to a value that contains an inert function cast that is an injection. Casing on the label of the active function cast's target type yields two cases: *cast-fun-id*$\star$ and *cast-fun-proj(-blame)*. Using $\bigcirc$ to represent the raw function types that we do not care about, the redex for *cast-fun-id*$\star$ is of shape $V \langle \bigcirc_\ell \Rightarrow \bigcirc_\star \rangle \langle \bigcirc_\star \Rightarrow \bigcirc_\star \rangle$. We can see that the first (inert) cast is an injection and the second (active) cast is an identity on $\star$. Consequently, we propagate the source of injection $\ell$ across and get: $V \langle \bigcirc_\ell \Rightarrow \bigcirc_\ell \rangle \langle \bigcirc_\ell \Rightarrow \bigcirc_\star \rangle$. On the other hand, in *cast-fun-proj* the second cast of the redex is a projection: $V \langle \bigcirc_{\ell_1} \Rightarrow \bigcirc_\star \rangle \langle \bigcirc_\star \Rightarrow \bigcirc_{\ell_4} \rangle$. We check whether $\ell_1 \preccurlyeq \ell_4$; if yes, we propagate $\ell_4$: $V \langle \bigcirc_{\ell_4} \Rightarrow \bigcirc_{\ell_4} \rangle \langle \bigcirc_{\ell_4} \Rightarrow \bigcirc_{\ell_4} \rangle$, otherwise we blame the projection (*cast-fun-proj-blame*). The rules for case (2), propagating PC labels in function casts, and the rules for reference casts follow the same basic idea.

The elimination rules for wrapped values are shown in Figure 6. Consider *if-cast-true*; the high level goal is to 1) reduce to the protected then-branch 2) convert the inert cast on the wrapped branch condition into a cast on the protected branch. We protect the then-branch $M$ with $\ell$ from the boolean constant as usual, because casts do not classify values. Then we insert a cast ($branch_c$ $A$ $c$) on the protected branch. The source and target types of the new cast are calculated by stamping the respective labels from $c$ (the cast on the branch condition) onto $A$ (the type of $M$) (8). To preserve types, we insert a PC cast around $M$ to adapt to the static PC of $M$, which is $\star$.

Next we discuss the *fun-cast* reduction. It eliminates a function proxy $V \langle c \rangle$ being applied to $W$. The high level idea is that we distribute the inert function cast into two casts: one on the domain side and the other on the co-domain side. The helper function *elim-fun-proxy* cases on the PC label of the target type, yielding two cases. If PC is concrete, nothing special is required to preserve types (9). Otherwise if PC is $\star$, we check if $pc \vee \ell_1 \preccurlyeq pc_1$ holds. If yes, we insert a PC cast and make sure that the static PC used to type the term before the co-domain cast equals to current dynamic PC $pc$. Otherwise, we blame cast $c$ because the labels on it, $pc_1$ and $\ell_1$, are ill-formed with respect to the current $pc$. The insertion of the check and the PC cast is guided by our type safety proof. An interesting observation about this check, equivalent to $pc \preccurlyeq pc_1 \wedge \ell_1 \preccurlyeq pc_1$, is that it is a direct analogue of the side condition on the typing rule of application in a fully static type system (7).

We obtain the big-step semantics in Figure 15 of the Appendix by a mechanical conversion from the small-step semantics. It has form $\mu \mid pc \vdash M \Downarrow V \mid \mu'$, relating term $M$ to value $V$. The big-step semantics only considers successful evaluations of $M$ and omits all the error cases because we use it to prove noninterference which is termination and error-insensitive. The protection term is not needed in a big-step semantics. Instead, we stamp PCs on values directly. Neither does PC cast appear in the big-step semantics, because value typing is agnostic about the PC (Lemma 3).

## 6 MECHANIZED META-THEORETICAL RESULTS

In this section we describe the proofs of four theorems: type safety, determinism, noninterference, and compilation preserves types. Everything is implemented in Agda and fully machine-checked, so we here we give an overview of the proofs and explain the main ideas.

### 6.1 Type Safety of $\lambda_{\mathtt{SEC}}^{\Rightarrow}$

We show that $\lambda_{\mathtt{SEC}}^{\Rightarrow}$ is type safe by proving progress (Theorem 2) and preservation (Theorem 4). We first define what it means for heap $\mu$ to be well-typed under context $\Sigma$:

Definition 1 (Heap typing). *$\Sigma \vdash \mu$ iff. for any $a$ that satisfies $\Sigma(a) = A$, there exists $V$ s.t lookup $\mu$ $a = V$ and $\emptyset; \Sigma; \mathtt{low}; \mathtt{low} \vdash V : A$.*

Note that we lookup $a$ in the half-heap that corresponds to the half-context. We prove that reference creation and assignment both preserve well-typedness of the heap.

Progress says that a well-typed $\lambda_{\mathtt{SEC}}^{\Rightarrow}$ term does not get stuck. The term is either be a value or an error, which does not reduce, or the term takes one reduction step further:

Theorem 2 (Progress). *Suppose $M$ is well-typed: $\emptyset; \Sigma; gc; pc \vdash M : A$ and the heap $\mu$ is also well-typed: $\Sigma \vdash \mu$. Then either (1) $M$ is a value or (2) $M$ is an error: $M = \mathtt{error}\ e$ for some $e$ or (3) $M$ can take a reduction step: $M \mid \mu \mid pc \longrightarrow M' \mid \mu'$ for some $M'$ and $\mu'$.*

Proof sketch. By induction on the typing derivation of $M$. In the NSU cases, case on $pc \preccurlyeq \ell$ (*⊢ref?*) and $pc \preccurlyeq \hat{\ell}$ (*⊢assign?*) respectively. Take one step by applying the success rule (*ref?-ok*, *assign?-ok*) if the NSU check passes or the failure rule (*ref?-fail*, *assign?-fail*) if it does not. □

The preservation proof is relatively straightforward. Preservation of parallel and single substitutions is proved by the usual approach [McBride 2005]. One major difference from GTLC, however, is that for a specific typing rule, in addition to types, we also require the PCs agree between the inner terms and the outer term.

One important observation is that with regard to typability, PCs do not matter for values, so we can arbitrarily replace them. This is why we have $pc$ annotations on $\lambda$s. The static PC used for type checking a $\lambda$'s body comes from the annotation and has nothing to do with the one that types the $\lambda$. We formalize this idea:

Lemma 3 (Value typing is agnostic about PCs). *If $\Gamma; \Sigma; gc; pc \vdash V : A$, then $\Gamma; \Sigma; gc'; pc' \vdash V : A$ , for any $gc', pc'$.*

Proof sketch. By induction on the typing derivation of $V$ and then inversion on the value. □

We now state the preservation theorem for both the small-step and big-step semantics.

Theorem 4 (Preservation). *Suppose $M$ is well-typed: $\emptyset; \Sigma; gc; pc \vdash M : A$ and the heap $\mu$ is also well-typed: $\Sigma \vdash \mu$. The static and dynamic PCs satisfy: $pc \lesssim gc$.*

*Small-step: If $M \mid \mu \mid pc \longrightarrow M' \mid \mu'$, there exists $\Sigma'$ s.t $\Sigma' \supseteq \Sigma$, $\emptyset; \Sigma'; gc; pc \vdash M' : A$, and $\Sigma' \vdash \mu'$.*

*Big-step: If $\mu \mid pc \vdash M \Downarrow V \mid \mu'$, there exists $\Sigma'$ s.t $\Sigma' \supseteq \Sigma$, $\emptyset; \Sigma'; gc; pc \vdash V : A$, and $\Sigma' \vdash \mu'$.*

Proof sketch. By induction on the reduction step and then inversion on the typing derivation of $M$. Use "single substitution preserves types" in $\beta$ and $\beta$-*let*. Use "reference creation preserves heap well-typedness" in *ref* and "assignment preserves heap well-typedness" in *assign*.

The proof for big-step is by induction on the big-step relation. Everything else is similar. □

### 6.2 Erasure and Determinism of Erased $\lambda_{\mathsf{SEC}}^{\Rightarrow}$

We define erased $\lambda_{\mathsf{SEC}}^{\Rightarrow}$ as the image of $\lambda_{\mathsf{SEC}}^{\Rightarrow}$ under the erasure function $\epsilon$. Erased $\lambda_{\mathsf{SEC}}^{\Rightarrow}$ is a subset of $\lambda_{\mathsf{SEC}}^{\Rightarrow}$. In this section, we prove that the big-step evaluation of erased $\lambda_{\mathsf{SEC}}^{\Rightarrow}$ is deterministic.

First we briefly talk about the erasure function $\epsilon$ on terms and heap (Figure 17 of the Appendix). As usual, high security constants and $\lambda$s are erased, replaced with ● (22) (23). For reasons mentioned in Section 3.2, we erase addresses unless both labels are low (21). As we have discussed in Section 2.2, we do not use type-guided classification, meaning that casts do not affect the security of values, thus they can be directly discarded (24). So are PC casts (25). The low-heap is erased point-wise (26) (27). The heap is erased by erasing the low-heap and ditching the high-heap (28).

The big-step semantics for erase $\lambda_{\mathsf{SEC}}^{\Rightarrow}$ is presented in Figure 18 of the Appendix. The $\mu$ is for low-heap only, because erasure discards the high-heap entirely. Basically, wherever constants, $\lambda$s, and addresses appear, their labels must be all low; otherwise they are erased into ●, which then follow the "-●" rules. In either $\Downarrow_\epsilon$-*ref?*-● or $\Downarrow_\epsilon$-*ref*-●, we skip the reference creation and erase the result, because it potentially produces an address that references the high-heap, which no longer exists.

Theorem 5 (Big-step evaluation of erased $\lambda_{\mathsf{SEC}}^{\Rightarrow}$ is deterministic). *If* $\mu \mid pc \vdash M \Downarrow_\epsilon V_1 \mid \mu_1$ *and* $\mu \mid pc \vdash M \Downarrow_\epsilon V_2 \mid \mu_2$, *then* $V_1 = V_2$ *and* $\mu_1 = \mu_2$.

Proof sketch. By induction on the first big-step and then inversion on the second. □

### 6.3 Noninterference

In this section, we assemble everything proved so far together and further prove noninterference.

The key to the noninterference proof is a simulation lemma (Lemma 7) between the unerased side and the erased side. One major challenge when proving this lemma is that we sometimes need to reason about side effects under high PC. Take $\Downarrow$-*if-true* (Figure 15 of the Appendix) for example, suppose $\ell = \mathtt{high}$, we know $\epsilon\ \mu \mid pc \vdash \epsilon\ L \Downarrow_\epsilon \bullet \mid \epsilon\ \mu_1$ and $\epsilon\ \mu_1 \mid \mathtt{high} \vdash \epsilon\ M \Downarrow_\epsilon \epsilon\ V \mid \epsilon\ \mu_2$ by induction hypotheses. We need to show $\epsilon\ \mu \mid pc \vdash \mathtt{if}\ (\epsilon\ L)\ A\ (\epsilon\ M)\ (\epsilon\ N) \Downarrow_\epsilon \bullet \mid \epsilon\ \mu_2$. We can construct the proof using rule $\Downarrow_\epsilon$-*if*-● (Figure 18 of the Appendix) if we know $\epsilon\ \mu_1 = \epsilon\ \mu_2$. This observation above brings about Lemma 6. It says that if we evaluate a term $M$ under high-PC, then the heaps before and after are related by erasure. Intuitively, it means that all side effects happening under a high-PC do not matter from a low-privileged observer's perspective.

Lemma 6 (Heaps are related by erasure under high PC). *Suppose* $\emptyset; \Sigma; gc; \mathtt{high} \vdash M : A$, $\Sigma \vdash \mu$, *and* $\mathtt{high} \precsim gc$. *If* $\mu \mid \mathtt{high} \vdash M \Downarrow V \mid \mu'$, *then* $\epsilon\ \mu = \epsilon\ \mu'$.

Proof sketch. By induction on the big-step and then inversion on the typing derivation. □

Lemma 7 (Simulation between original and erased $\lambda_{\mathsf{SEC}}^{\Rightarrow}$). *Suppose* $\emptyset; \Sigma; gc; pc \vdash M : A$, $\Sigma \vdash \mu$, *and* $pc \precsim gc$. *If* $\mu \mid pc \vdash M \Downarrow V \mid \mu'$, *then* $\epsilon\ \mu \mid pc \vdash \epsilon\ M \Downarrow_\epsilon \epsilon\ V \mid \epsilon\ \mu'$.

Proof sketch. By induction on the big-step relation and then inversion on the typing derivation of $M$. Case on $\ell$ in $\Downarrow$-*app*, $\Downarrow$-*if*, and $\Downarrow$-*if-cast*. Use Lemma 6 when $\ell = \mathtt{high}$. □

We state noninterference in Theorem 8. The input is modeled as a free variable $x$ and the output is the evaluation result of the $\lambda_{\mathsf{SEC}}^{\Rightarrow}$ term $M$. The typing judgment of $M$ says that the input is a high-security boolean constant while the output is a low-security boolean constant. Both PCs, static and dynamic, are originally $\mathtt{low}$ and the heap is empty. If we run $M$ with two values, $(\$\ b_1)_{\mathtt{high}}$ and $(\$\ b_2)_{\mathtt{high}}$, which potentially carry different user-input data, Theorem 8 tells us that the observable computation results (values) of the two executions, $V_1$ and $V_2$, are equal.

Theorem 8 (Noninterference). *If $M$ is well-typed:* $(x{:}\mathtt{Bool}_{\mathtt{high}}); \emptyset; \mathtt{low}; \mathtt{low} \vdash M : \mathtt{Bool}_{\mathtt{low}}$ *and*

$$\emptyset \mid \mathtt{low} \vdash M[x := (\$\ b_1)_{\mathtt{high}}] \Downarrow V_1 \mid \mu_1 \quad \text{and} \quad \emptyset \mid \mathtt{low} \vdash M[x := (\$\ b_2)_{\mathtt{high}}] \Downarrow V_2 \mid \mu_2$$

*then* $V_1 = V_2$.

Proof. Applying the simulation lemma (Lemma 7) on the premises respectively, we get:

$$\emptyset \mid \mathtt{low} \vdash \epsilon\ M[x := \bullet] \Downarrow \epsilon\ V_1 \mid \epsilon\ \mu_1 \quad \text{and} \quad \emptyset \mid \mathtt{low} \vdash \epsilon\ M[x := \bullet] \Downarrow \epsilon\ V_2 \mid \epsilon\ \mu_2$$

Note that after erasure, the left hand sides of big-step in the above become the same. We apply the determinism theorem (Theorem 5) to obtain $\epsilon\ V_1 = \epsilon\ V_2$. We know $\vdash V_i : \mathtt{Bool}_{\mathtt{low}}, i \in \{1, 2\}$ because big-step preserves types (Theorem 4). Consequently, we know $V_i = (\$\ b_i)_{\mathtt{low}}, i \in \{1, 2\}$ due to the canonical form of constants. So we have $\epsilon\ (\$\ b_1)_{\mathtt{low}} = \epsilon\ (\$\ b_2)_{\mathtt{low}}$. By the definition of erasure $\epsilon$ (Figure 17 of the Appendix), we have $(\$\ b_1)_{\mathtt{low}} = (\$\ b_2)_{\mathtt{low}}$, which is equivalent to $V_1 = V_2$. □

## 6.4 Compilation from $\lambda^{\star}_{\mathtt{SEC}}$ to $\lambda^{\Rightarrow}_{\mathtt{SEC}}$ Preserves Types

Finally, we connect the surface language and its intermediate representation by proving that compiling from $\lambda^{\star}_{\mathtt{SEC}}$ to $\lambda^{\Rightarrow}_{\mathtt{SEC}}$ preserves types.

Lemma 9. *If* $\Gamma; gc \vdash M : A$*, then* $\Gamma; \emptyset; gc; pc \vdash \mathcal{C}\ M : A$ *for any* $pc$.

Proof sketch. By induction on the typing derivation of $M$ and follow the definition of $\mathcal{C}$. □

Theorem 10 (Compilation preserves types). *If* $\Gamma; gc \vdash M : A$*, then* $\Gamma; \emptyset; gc; \mathtt{low} \vdash \mathcal{C}\ M : A$.

Proof. By instantiating $pc = \mathtt{low}$ in Lemma 9. □

# 7 CONCLUSION

We have presented an information-flow control language, named $\lambda^{\star}_{\mathtt{SEC}}$, that is gradual in the sense that the programmer decides whether the IFC occurs statically or dynamically in different regions of their program. This paper presents the first mechanized proof of noninterference for such a language. The prior mechanized proofs of noninterference by Stefan et al. [2017] and Xiang and Chong [2021] were for languages with dynamic control of information-flow, but not for static or gradual control. Compared to pen-and-paper proofs of noninterference for gradually-typed information-flow languages, our proof is most similar to the flawed proof of Fennell and Thiemann [2013] that also uses the erasure approach; the main differences are that we use a big-step semantics instead of small-step and we use a split heap to fix how erasure handles addresses. Toro et al. [2018] and Azevedo de Amorim et al. [2020] also develop pen-and-paper proofs of noninterference for gradually-typed languages, and we are not aware of any flaws in those proofs, but the proof techniques that they use are less amenable to mechanization.

Our language $\lambda^{\star}_{\mathtt{SEC}}$ is based on the GLIO language of Azevedo de Amorim et al. [2020], which satisfies both noninterference and the dynamic gradual guarantee. (The $\mathrm{GSL}_{\mathrm{Ref}}$ language of Toro et al. [2018] satisfies noninterference but not the dynamic gradual guarantee.) Azevedo de Amorim et al. [2020] choose to define GLIO via denotational semantics, which differs from the rest of the literature on gradual typing, making their results difficult to build on by other researchers. In this paper we contribute a traditional semantics for $\lambda^{\star}_{\mathtt{SEC}}$, whose semantics is defined by 1) compilation to a cast calculus and 2) a reduction semantics for the cast calculus.

## APPENDIX

$$\boxed{g_1 <: g_2,\ S <: T,\ \text{and}\ A <: B}$$

$$\textit{<:-}\star\ \frac{}{\star <: \star} \qquad \textit{<:-}\ell\ \frac{\ell_1 \preccurlyeq \ell_2}{\ell_1 <: \ell_2} \qquad \textit{<:-}\iota\ \frac{}{\iota <: \iota} \qquad \textit{<:-ref}\ \frac{A <: B \quad B <: A}{\texttt{Ref}\ A <: \texttt{Ref}\ B}$$

$$\textit{<:-fun}\ \frac{gc_2 <: gc_1 \quad C <: A \quad B <: D}{A \xrightarrow{gc_1} B <: C \xrightarrow{gc_2} D} \qquad \textit{<:-}\tau\ \frac{g_1 <: g_2 \quad S <: T}{S_{g_1} <: T_{g_2}}$$

Fig. 7. Subtyping of labels and types

$$\boxed{g_1 \sim g_2,\ S \sim T,\ \text{and}\ A \sim B}$$

$$\star\!\sim\ \frac{}{\star \sim g} \qquad \sim\!\star\ \frac{}{g \sim \star} \qquad \ell\!\sim\ \frac{}{\ell \sim \ell} \qquad \sim\!\textit{-}\iota\ \frac{}{\iota \sim \iota} \qquad \sim\!\textit{-ref}\ \frac{A \sim B}{\texttt{Ref}\ A \sim \texttt{Ref}\ B}$$

$$\sim\!\textit{-fun}\ \frac{gc_1 \sim gc_2 \quad A \sim C \quad B \sim D}{A \xrightarrow{gc_1} B \sim C \xrightarrow{gc_2} D} \qquad \sim\!\textit{-}\tau\ \frac{g_1 \sim g_2 \quad S \sim T}{S_{g_1} \sim T_{g_2}}$$

Fig. 8. Consistency for labels and types

$$\boxed{g_1 \lesssim g_2,\ S \lesssim T,\ \text{and}\ A \lesssim B}$$

$$\lesssim\!\star\ \frac{}{g \lesssim \star} \qquad \star\!\lesssim\ \frac{}{\star \lesssim g} \qquad \lesssim\!\textit{-}\ell\ \frac{\ell_1 \preccurlyeq \ell_2}{\ell_1 \lesssim \ell_2} \qquad \lesssim\!\textit{-}\iota\ \frac{}{\iota \lesssim \iota} \qquad \lesssim\!\textit{-ref}\ \frac{A \lesssim B \quad B \lesssim A}{\texttt{Ref}\ A \lesssim \texttt{Ref}\ B}$$

$$\lesssim\!\textit{-fun}\ \frac{gc_2 \lesssim gc_1 \quad C \lesssim A \quad B \lesssim D}{A \xrightarrow{gc_1} B \lesssim C \xrightarrow{gc_2} D} \qquad \lesssim\!\textit{-}\tau\ \frac{g_1 \lesssim g_2 \quad S \lesssim T}{S_{g_1} \lesssim T_{g_2}}$$

Fig. 9. Consistent subtyping for labels and types

$$
\begin{aligned}
\ell \sqcap \ell &= \ell \\
\star \sqcap g &= g \\
g \sqcap \star &= g \\
\iota \sqcap \iota &= \iota \\
(\texttt{Ref}\ A) \sqcap (\texttt{Ref}\ B) &= \texttt{Ref}\ A' \text{ where } A' = A \sqcap B \\
(A \xrightarrow{gc_1} B) \sqcap (C \xrightarrow{gc_2} D) &= A' \xrightarrow{gc} B' \\
&\quad \text{where } gc = gc_1 \sqcap gc_2,\ A' = A \sqcap C, \text{ and } B' = B \sqcap D \\
S_{g_1} \sqcap T_{g_2} &= T'_g \\
&\quad \text{where } T' = S \sqcap T \text{ and } g = g_1 \sqcap g_2 \\
&\quad \text{(-}\sqcap\text{- is undefined otherwise)}
\end{aligned}
$$

$$
\begin{aligned}
\ell_1 \mathbin{\widetilde{\curlyvee}} \ell_2 &= \ell_1 \vee \ell_2 \\
- \mathbin{\widetilde{\curlyvee}} \star &= \star \\
\star \mathbin{\widetilde{\curlyvee}} - &= \star \\
\iota \mathbin{\widetilde{\vee}} \iota &= \iota \\
(\texttt{Ref}\ A) \mathbin{\widetilde{\vee}} (\texttt{Ref}\ B) &= \texttt{Ref}\ C \text{ where } C = A \sqcap B \\
(A \xrightarrow{gc_1} B) \mathbin{\widetilde{\vee}} (C \xrightarrow{gc_2} D) &= A' \xrightarrow{gc_1 \mathbin{\widetilde{\curlywedge}} gc_2} B' \\
&\quad \text{where } A' = A \mathbin{\widetilde{\wedge}} C \text{ and } B' = B \mathbin{\widetilde{\vee}} D \\
S_{g_1} \mathbin{\widetilde{\vee}} T_{g_2} &= T'_{g_1 \mathbin{\widetilde{\curlyvee}} g_2} \text{ where } T' = S \mathbin{\widetilde{\vee}} T \\
&\quad \text{(-}\mathbin{\widetilde{\vee}}\text{- is undefined otherwise)}
\end{aligned}
$$

$$
\begin{aligned}
\ell_1 \mathbin{\widetilde{\curlywedge}} \ell_2 &= \ell_1 \wedge \ell_2 \\
- \mathbin{\widetilde{\curlywedge}} \star &= \star \\
\star \mathbin{\widetilde{\curlywedge}} - &= \star \\
\iota \mathbin{\widetilde{\wedge}} \iota &= \iota \\
(\texttt{Ref}\ A) \mathbin{\widetilde{\wedge}} (\texttt{Ref}\ B) &= \texttt{Ref}\ C \text{ where } C = A \sqcap B \\
(A \xrightarrow{gc_1} B) \mathbin{\widetilde{\wedge}} (C \xrightarrow{gc_2} D) &= A' \xrightarrow{gc_1 \mathbin{\widetilde{\curlyvee}} gc_2} B' \\
&\quad \text{where } A' = A \mathbin{\widetilde{\vee}} C \text{ and } B' = B \mathbin{\widetilde{\wedge}} D \\
S_{g_1} \mathbin{\widetilde{\wedge}} T_{g_2} &= T'_{g_1 \mathbin{\widetilde{\curlywedge}} g_2} \text{ where } T' = S \mathbin{\widetilde{\wedge}} T \\
&\quad \text{(-}\mathbin{\widetilde{\wedge}}\text{- is undefined otherwise)}
\end{aligned}
$$

Fig. 10. Operators for gradual labels and types: gradual meet (-⊓-), consistent join ($-\mathbin{\widetilde{\curlyvee}}-$ for labels and $-\mathbin{\widetilde{\vee}}-$ for types), and consistent meet ($-\mathbin{\widetilde{\curlywedge}}-$ for labels and $-\mathbin{\widetilde{\wedge}}-$ for types)

| | |
|---|---|
| $- \leftharpoonup - : (g_1\ g_2 : \mathit{Label}) \to \mathit{Label}$ | , where $g_1 \lesssim g_2$ |
| $- \leftharpoonup - : (S\ T : \mathit{RawType}) \to \mathit{RawType}$ | , where $S \lesssim T$ |
| $- \leftharpoonup - : (A\ B : \mathit{Type}) \to \mathit{Type}$ | , where $A \lesssim B$ |

$$
\begin{aligned}
- \leftharpoonup \star &= \star \\
\star \leftharpoonup g &= g \\
\ell_1 \leftharpoonup \ell_2 &= \ell_1 \\
\iota \leftharpoonup \iota &= \iota \\
\mathtt{Ref}\ A \leftharpoonup \mathtt{Ref}\ B &= \mathtt{Ref}\ B \\
A \xrightarrow{gc_1} B \leftharpoonup C \xrightarrow{gc_2} D &= A' \xrightarrow{gc} B' \\
&\quad \text{where } gc = gc_2 \leftharpoonup gc_1, \\
&\quad\quad A' = C \leftharpoonup A, \text{ and } B' = B \leftharpoonup D \\
S_{g_1} \leftharpoonup T_{g_2} &= T'_g \\
&\quad \text{where } T' = S \leftharpoonup T \text{ and } g = g_1 \leftharpoonup g_2
\end{aligned}
$$

Fig. 11. Merge operators for labels and types

$$\boxed{C\ M \rightsquigarrow M'}$$

$$C\ (\$\ k)_\ell \rightsquigarrow (\$\ k)_\ell \tag{11}$$

$$C\ x \rightsquigarrow x \tag{12}$$

$$C\ (\lambda^{pc} x{:}A.\ N)_\ell \rightsquigarrow (\lambda^{pc} x{:}A.\ N')_\ell \tag{13}$$
$$\text{where } N \rightsquigarrow N'$$

$$C\ (L\ M)^p \rightsquigarrow L' \langle c_1 \rangle\ M' \langle c_2 \rangle \tag{14}$$
$$\text{where}$$
$$L \rightsquigarrow L', M \rightsquigarrow M', C = A' \leftharpoondown A, g_1 = gc \leftharpoondown gc', g_2 = g \leftharpoondown gc'$$
$$c_1 = (A \xrightarrow{gc'} B)_g \Rightarrow^p (A \xrightarrow{g_1 \mathbin{\widetilde{\curlyvee}} g_2} B)_g, c_2 = A' \Rightarrow^p C$$
$$\Gamma; gc \vdash L : (A \xrightarrow{gc'} B)_g, \Gamma; gc \vdash M : A'$$

$$C\ (\mathtt{if}\ L\ \mathtt{then}\ M\ \mathtt{else}\ N)^p \rightsquigarrow \mathtt{if}\ L'\ C\ M' \langle c_1 \rangle\ N' \langle c_2 \rangle \tag{15}$$
$$\text{where}$$
$$L \rightsquigarrow L', M \rightsquigarrow M', N \rightsquigarrow N', A' = A \leftharpoondown C, B' = B \leftharpoondown C$$
$$c_1 = A \Rightarrow^p A', c_2 = B \Rightarrow^p B'$$
$$\Gamma; gc \vdash L : \mathtt{Bool}_g, \Gamma; gc \mathbin{\widetilde{\curlyvee}} g \vdash M : A, \Gamma; gc \mathbin{\widetilde{\curlyvee}} g \vdash N : B$$
$$C = A \mathbin{\widetilde{\curlyvee}} B \text{ (therefore } A \lesssim C, B \lesssim C)$$

$$C\ (M : A)^p \rightsquigarrow M' \langle A' \Rightarrow^p B \rangle \tag{16}$$
$$\text{where } M \rightsquigarrow M', B = A' \leftharpoondown A, \Gamma; gc \vdash M : A'$$

$$C\ (\mathtt{let}\ x = M\ \mathtt{in}\ N) \rightsquigarrow \mathtt{let}\ x = M\ \mathtt{in}\ N \tag{17}$$
$$\text{where } M \rightsquigarrow M', N \rightsquigarrow N'$$

$$C\ (\mathtt{ref}\ \ell\ M)^p \rightsquigarrow \begin{cases} \mathtt{ref}\ \ \ell\ M' \langle T_g \Rightarrow^p A \rangle & \text{, if } gc \text{ is concrete} \\ \mathtt{ref}^?\ \ell\ M' \langle T_g \Rightarrow^p A \rangle & \text{, if } gc = \star \end{cases} \tag{18}$$
$$\text{where } M \rightsquigarrow M', A = T_g \leftharpoondown T_\ell, \Gamma; gc \vdash M : T_g$$

$$C\ (!\ M) \rightsquigarrow !\ M \tag{19}$$
$$\text{where } M \rightsquigarrow M'$$

$$C\ (L := M)^p \rightsquigarrow \begin{cases} L' \langle c_1 \rangle \mathrel{:=} M' \langle c_2 \rangle & \text{, if } gc \text{ and } \hat{g} \text{ are both concrete} \\ L' \langle c_1 \rangle \mathrel{:=^?} M' \langle c_2 \rangle & \text{, if } gc = \star \text{ or } \hat{g} = \star \end{cases} \tag{20}$$
$$\text{where}$$
$$L \rightsquigarrow L', M \rightsquigarrow M', B = A \leftharpoondown T_{\hat{g}}, g' = g \leftharpoondown \hat{g}$$
$$c_1 = (\mathtt{Ref}\ T_{\hat{g}})_g \Rightarrow^p (\mathtt{Ref}\ T_{\hat{g}})_{g'}, c_2 = A \Rightarrow^p B$$
$$\Gamma; gc \vdash L : (\mathtt{Ref}\ T_{\hat{g}})_g, \Gamma; gc \vdash M : A$$

Fig. 12. Compilation from surface language $\lambda^{\star}_{\mathsf{SEC}}$ to cast calculus $\lambda^{\Rightarrow}_{\mathsf{SEC}}$

$$\boxed{\textbf{Active}\ g_1 \Rightarrow g_2 \text{ and } \textbf{Active}\ A \Rightarrow B}$$

$$\textit{A-label-id}\star\ \frac{}{\textbf{Active}\ \star \Rightarrow \star} \qquad \textit{A-label-proj}\ \frac{}{\textbf{Active}\ \star \Rightarrow \ell}$$

$$\textit{A-base-id}\ \frac{}{\textbf{Active}\ \iota_g \Rightarrow \iota_g} \qquad \textit{A-base-proj}\ \frac{}{\textbf{Active}\ \iota_\star \Rightarrow \iota_\ell}$$

$$\textit{A-fun}\ \frac{\textbf{Active}\ g_1 \Rightarrow g_2}{\textbf{Active}\ (A \xrightarrow{gc_1} B)_{g_1} \Rightarrow (C \xrightarrow{gc_2} D)_{g_2}} \qquad \textit{A-fun-pc}\ \frac{\textbf{Active}\ gc_1 \Rightarrow gc_2 \quad \textbf{Inert}\ g_1 \Rightarrow g_2}{\textbf{Active}\ (A \xrightarrow{gc_1} B)_{g_1} \Rightarrow (C \xrightarrow{gc_2} D)_{g_2}}$$

$$\textit{A-ref}\ \frac{\textbf{Active}\ g_1 \Rightarrow g_2}{\textbf{Active}\ (\texttt{Ref}\ A)_{g_1} \Rightarrow (\texttt{Ref}\ B)_{g_2}} \qquad \textit{A-ref-ref}\ \frac{\textbf{Active}\ \hat{g}_1 \Rightarrow \hat{g}_2 \quad \textbf{Inert}\ g_1 \Rightarrow g_2}{\textbf{Active}\ (\texttt{Ref}\ S_{\hat{g}_1})_{g_1} \Rightarrow (\texttt{Ref}\ T_{\hat{g}_2})_{g_2}}$$

$$\boxed{\textbf{Inert}\ g_1 \Rightarrow g_2 \text{ and } \textbf{Inert}\ A \Rightarrow B}$$

$$\textit{I-label}\ \frac{}{\textbf{Inert}\ \ell \Rightarrow g} \qquad \textit{I-base-inj}\ \frac{}{\textbf{Inert}\ \iota_\ell \Rightarrow \iota_\star}$$

$$\textit{I-fun}\ \frac{\textbf{Inert}\ gc_1 \Rightarrow gc_2 \quad \textbf{Inert}\ g_1 \Rightarrow g_2}{\textbf{Inert}\ (A \xrightarrow{gc_1} B)_{g_1} \Rightarrow (C \xrightarrow{gc_2} D)_{g_2}} \qquad \textit{I-ref}\ \frac{\textbf{Inert}\ \hat{g}_1 \Rightarrow \hat{g}_2 \quad \textbf{Inert}\ g_1 \Rightarrow g_2}{\textbf{Inert}\ (\texttt{Ref}\ S_{\hat{g}_1})_{g_1} \Rightarrow (\texttt{Ref}\ T_{\hat{g}_2})_{g_2}}$$

Fig. 13. Active casts and inert casts

$$\boxed{\textbf{Cast}\ V\ ,\ c \rightsquigarrow M}$$

$$\textit{cast-base-id}\ \frac{}{\textbf{Cast}\ V\ ,\ \iota_g \Rightarrow^{p} \iota_g \rightsquigarrow V} \qquad \textit{cast-base-proj}\ \frac{\ell_1 \preccurlyeq \ell_2}{\textbf{Cast}\ V\ \langle\ \iota_{\ell_1} \Rightarrow^{p} \iota_\star\ \rangle\ ,\ \iota_\star \Rightarrow^{q} \iota_{\ell_2} \rightsquigarrow V}$$

$$\textit{cast-base-proj-blame}\ \frac{\ell_1 \not\preccurlyeq \ell_2}{\textbf{Cast}\ V\ \langle\ \iota_{\ell_1} \Rightarrow^{p} \iota_\star\ \rangle\ ,\ \iota_\star \Rightarrow^{q} \iota_{\ell_2} \rightsquigarrow \texttt{error}\ \texttt{blame}^{q}}$$

$$\textit{cast-fun-id}\star\ \frac{}{\begin{array}{l}\textbf{Cast}\ V\ \langle\ (A_1 \xrightarrow{gc_1} B_1)_\ell \Rightarrow^{p} (A_2 \xrightarrow{gc_2} B_2)_\star\ \rangle\ ,\ (A_3 \xrightarrow{gc_3} B_3)_\star \Rightarrow^{q} (A_4 \xrightarrow{gc_4} B_4)_\star \rightsquigarrow \\ \quad V\ \langle\ (A_1 \xrightarrow{gc_1} B_1)_\ell \Rightarrow^{p} (A_2 \xrightarrow{gc_2} B_2)_\ell\ \rangle\ \langle\ (A_3 \xrightarrow{gc_3} B_3)_\ell \Rightarrow^{q} (A_4 \xrightarrow{gc_4} B_4)_\star\ \rangle\end{array}}$$

$$\textit{cast-fun-proj}\ \frac{\ell_1 \preccurlyeq \ell_4}{\begin{array}{l}\textbf{Cast}\ V\ \langle\ (A_1 \xrightarrow{gc_1} B_1)_{\ell_1} \Rightarrow^{p} (A_2 \xrightarrow{gc_2} B_2)_\star\ \rangle\ ,\ (A_3 \xrightarrow{gc_3} B_3)_\star \Rightarrow^{q} (A_4 \xrightarrow{gc_4} B_4)_{\ell_4} \rightsquigarrow \\ \quad V\ \langle\ (A_1 \xrightarrow{gc_1} B_1)_{\ell_4} \Rightarrow^{p} (A_2 \xrightarrow{gc_2} B_2)_{\ell_4}\ \rangle\ \langle\ (A_3 \xrightarrow{gc_3} B_3)_{\ell_4} \Rightarrow^{q} (A_4 \xrightarrow{gc_4} B_4)_{\ell_4}\ \rangle\end{array}}$$

$$\textit{cast-fun-proj-blame}\ \frac{\ell_1 \not\preccurlyeq \ell_4}{\textbf{Cast}\ V\ \langle\ (A_1 \xrightarrow{gc_1} B_1)_{\ell_1} \Rightarrow^{p} (A_2 \xrightarrow{gc_2} B_2)_\star\ \rangle\ ,\ (A_3 \xrightarrow{gc_3} B_3)_\star \Rightarrow^{q} (A_4 \xrightarrow{gc_4} B_4)_{\ell_4} \rightsquigarrow \texttt{error}\ \texttt{blame}^{q}}$$

$$\textit{cast-fun-pc-id}\star\ \frac{}{\begin{array}{l}\textbf{Cast}\ V\ \langle\ (A_1 \xrightarrow{pc} B_1)_{g_1} \Rightarrow^{p} (A_2 \xrightarrow{\star} B_2)_{g_2}\ \rangle\ ,\ (A_3 \xrightarrow{\star} B_3)_{\ell_3} \Rightarrow^{q} (A_4 \xrightarrow{\star} B_4)_{g_4} \rightsquigarrow \\ \quad V\ \langle\ (A_1 \xrightarrow{pc} B_1)_{g_1} \Rightarrow^{p} (A_2 \xrightarrow{pc} B_2)_{g_2}\ \rangle\ \langle\ (A_3 \xrightarrow{pc} B_3)_{\ell_3} \Rightarrow^{q} (A_4 \xrightarrow{\star} B_4)_{g_4}\ \rangle\end{array}}$$

$$\textit{cast-fun-pc-proj}\ \frac{pc_4 \preccurlyeq pc_1}{\begin{array}{l}\textbf{Cast}\ V\ \langle\ (A_1 \xrightarrow{pc_1} B_1)_{g_1} \Rightarrow^{p} (A_2 \xrightarrow{\star} B_2)_{g_2}\ \rangle\ ,\ (A_3 \xrightarrow{\star} B_3)_{\ell_3} \Rightarrow^{q} (A_4 \xrightarrow{pc_4} B_4)_{g_4} \rightsquigarrow \\ \quad V\ \langle\ (A_1 \xrightarrow{pc_4} B_1)_{g_1} \Rightarrow^{p} (A_2 \xrightarrow{pc_4} B_2)_{g_2}\ \rangle\ \langle\ (A_3 \xrightarrow{pc_4} B_3)_{\ell_3} \Rightarrow^{q} (A_4 \xrightarrow{pc_4} B_4)_{g_4}\ \rangle\end{array}}$$

$$\textit{cast-fun-pc-proj-blame}\ \frac{pc_4 \not\preccurlyeq pc_1}{\textbf{Cast}\ V\ \langle\ (A_1 \xrightarrow{pc_1} B_1)_{g_1} \Rightarrow^{p} (A_2 \xrightarrow{\star} B_2)_{g_2}\ \rangle\ ,\ (A_3 \xrightarrow{\star} B_3)_{\ell_3} \Rightarrow^{q} (A_4 \xrightarrow{pc_4} B_4)_{g_4} \rightsquigarrow \texttt{error}\ \texttt{blame}^{q}}$$

$$\textit{cast-ref-id}\star\ \frac{}{\begin{array}{l}\textbf{Cast}\ V\ \langle\ (\texttt{Ref}\ A)_\ell \Rightarrow^{p} (\texttt{Ref}\ B)_\star\ \rangle\ ,\ (\texttt{Ref}\ C)_\star \Rightarrow^{q} (\texttt{Ref}\ D)_\star \rightsquigarrow \\ \quad V\ \langle\ (\texttt{Ref}\ A)_\ell \Rightarrow^{p} (\texttt{Ref}\ B)_\ell\ \rangle\ \langle\ (\texttt{Ref}\ C)_\ell \Rightarrow^{q} (\texttt{Ref}\ D)_\star\ \rangle\end{array}}$$

$$\textit{cast-ref-proj}\ \frac{\ell_1 \preccurlyeq \ell_4}{\begin{array}{l}\textbf{Cast}\ V\ \langle\ (\texttt{Ref}\ A)_{\ell_1} \Rightarrow^{p} (\texttt{Ref}\ B)_\star\ \rangle\ ,\ (\texttt{Ref}\ C)_\star \Rightarrow^{q} (\texttt{Ref}\ D)_{\ell_4} \rightsquigarrow \\ \quad V\ \langle\ (\texttt{Ref}\ A)_{\ell_4} \Rightarrow^{p} (\texttt{Ref}\ B)_{\ell_4}\ \rangle\ \langle\ (\texttt{Ref}\ C)_{\ell_4} \Rightarrow^{q} (\texttt{Ref}\ D)_{\ell_4}\ \rangle\end{array}}$$

$$\textit{cast-ref-proj-blame}\ \frac{\ell_1 \not\preccurlyeq \ell_4}{\textbf{Cast}\ V\ \langle\ (\texttt{Ref}\ A)_{\ell_1} \Rightarrow^{p} (\texttt{Ref}\ B)_\star\ \rangle\ ,\ (\texttt{Ref}\ C)_\star \Rightarrow^{q} (\texttt{Ref}\ D)_{\ell_4} \rightsquigarrow \texttt{error}\ \texttt{blame}^{q}}$$

$$\textit{cast-ref-ref-id}\star\ \frac{}{\begin{array}{l}\textbf{Cast}\ V\ \langle\ (\texttt{Ref}\ (T_1)_{\hat\ell})_{g_1} \Rightarrow^{p} (\texttt{Ref}\ (T_2)_\star)_{g_2}\ \rangle\ ,\ (\texttt{Ref}\ (T_3)_\star)_{\ell_3} \Rightarrow^{q} (\texttt{Ref}\ (T_4)_\star)_{g_4} \rightsquigarrow \\ \quad V\ \langle\ (\texttt{Ref}\ (T_1)_{\hat\ell})_{g_1} \Rightarrow^{p} (\texttt{Ref}\ (T_2)_{\hat\ell})_{g_2}\ \rangle\ \langle\ (\texttt{Ref}\ (T_3)_{\hat\ell})_{\ell_3} \Rightarrow^{q} (\texttt{Ref}\ (T_4)_\star)_{g_4}\ \rangle\end{array}}$$

$$\textit{cast-ref-ref-proj}\ \frac{\hat\ell_1 = \hat\ell_4}{\begin{array}{l}\textbf{Cast}\ V\ \langle\ (\texttt{Ref}\ (T_1)_{\hat\ell_1})_{g_1} \Rightarrow^{p} (\texttt{Ref}\ (T_2)_\star)_{g_2}\ \rangle\ ,\ (\texttt{Ref}\ (T_3)_\star)_{\ell_3} \Rightarrow^{q} (\texttt{Ref}\ (T_4)_{\hat\ell_4})_{g_4} \rightsquigarrow \\ \quad V\ \langle\ (\texttt{Ref}\ (T_1)_{\hat\ell_4})_{g_1} \Rightarrow^{p} (\texttt{Ref}\ (T_2)_{\hat\ell_4})_{g_2}\ \rangle\ \langle\ (\texttt{Ref}\ (T_3)_{\hat\ell_4})_{\ell_3} \Rightarrow^{q} (\texttt{Ref}\ (T_4)_{\hat\ell_4})_{g_4}\ \rangle\end{array}}$$

$$\textit{cast-ref-ref-proj-blame}\ \frac{\hat\ell_1 \neq \hat\ell_4}{\textbf{Cast}\ V\ \langle\ (\texttt{Ref}\ (T_1)_{\hat\ell_1})_{g_1} \Rightarrow^{p} (\texttt{Ref}\ (T_2)_\star)_{g_2}\ \rangle\ ,\ (\texttt{Ref}\ (T_3)_\star)_{\ell_3} \Rightarrow^{q} (\texttt{Ref}\ (T_4)_{\hat\ell_4})_{g_4} \rightsquigarrow \texttt{error}\ \texttt{blame}^{q}}$$

Fig. 14. Application rules for active casts

$$\boxed{\mu \mid pc \vdash M \Downarrow V \mid \mu'}$$

$$\Downarrow\textit{-val}\ \frac{}{\mu \mid pc \vdash V \Downarrow V \mid \mu} \qquad \Downarrow\textit{-app}\ \frac{\begin{array}{c}\mu \mid pc \vdash L \Downarrow (\lambda^{pc'} x{:}A.\ N)_\ell \mid \mu_1 \\ \mu_1 \mid pc \vdash M \Downarrow V \mid \mu_2 \\ \mu_2 \mid pc \curlyvee \ell \vdash N[x := V] \Downarrow W \mid \mu_3\end{array}}{\mu \mid pc \vdash L\ M \Downarrow W \curlyvee \ell \mid \mu_3}$$

$$\Downarrow\textit{-if-true}\ \frac{\begin{array}{c}\mu \mid pc \vdash L \Downarrow (\$\ \mathtt{true})_\ell \mid \mu_1 \\ \mu_1 \mid pc \curlyvee \ell \vdash M \Downarrow V \mid \mu_2\end{array}}{\mu \mid pc \vdash \mathtt{if}\ L\ A\ M\ N \Downarrow V \curlyvee \ell \mid \mu_2} \qquad \Downarrow\textit{-if-false}\ \frac{\begin{array}{c}\mu \mid pc \vdash L \Downarrow (\$\ \mathtt{false})_\ell \mid \mu_1 \\ \mu_1 \mid pc \curlyvee \ell \vdash N \Downarrow V \mid \mu_2\end{array}}{\mu \mid pc \vdash \mathtt{if}\ L\ A\ M\ N \Downarrow V \curlyvee \ell \mid \mu_2}$$

$$\Downarrow\textit{-let}\ \frac{\begin{array}{c}\mu \mid pc \vdash M \Downarrow V \mid \mu_1 \\ \mu_1 \mid pc \vdash N[x := V] \Downarrow W \mid \mu_2\end{array}}{\mu \mid pc \vdash \mathtt{let}\ x = M\ \mathtt{in}\ N \Downarrow W \mid \mu_2} \qquad \Downarrow\textit{-deref}\ \frac{\begin{array}{c}\mu \mid pc \vdash M \Downarrow (\mathtt{addr}\ a)_\ell \mid \mu_1 \\ \textit{lookup}\ \mu_1\ a = V\end{array}}{\mu \mid pc \vdash\ !\ M \Downarrow V \curlyvee \hat{\ell} \curlyvee \ell \mid \mu_1}\text{, where } a = n_{\hat{\ell}}$$

$$\Downarrow\textit{-ref?}\ \frac{\mu \mid pc \vdash M \Downarrow V \mid \mu_1 \qquad a = n_\ell\ \textbf{FreshIn}\ \mu_1 \qquad pc \preccurlyeq \ell}{\mu \mid pc \vdash \mathtt{ref}^?\ \ell\ M \Downarrow (\mathtt{addr}\ a)_{\mathtt{low}} \mid \textit{cons}\ a\ V\ \mu_1}$$

$$\Downarrow\textit{-ref}\ \frac{\mu \mid pc \vdash M \Downarrow V \mid \mu_1 \qquad a = n_\ell\ \textbf{FreshIn}\ \mu_1}{\mu \mid pc \vdash \mathtt{ref}\ \ell\ M \Downarrow (\mathtt{addr}\ a)_{\mathtt{low}} \mid \textit{cons}\ a\ V\ \mu_1}$$

$$\Downarrow\textit{-assign?}\ \frac{\begin{array}{c}\mu \mid pc \vdash L \Downarrow (\mathtt{addr}\ a)_\ell \mid \mu_1 \\ \mu_1 \mid pc \vdash M \Downarrow V \mid \mu_2 \qquad pc \preccurlyeq \hat{\ell}\end{array}}{\mu \mid pc \vdash L :=^? M \Downarrow (\$\ \mathtt{unit})_{\mathtt{low}} \mid \textit{cons}\ a\ V\ \mu_2}\text{, where } a = n_{\hat{\ell}}$$

$$\Downarrow\textit{-assign}\ \frac{\begin{array}{c}\mu \mid pc \vdash L \Downarrow (\mathtt{addr}\ a)_\ell \mid \mu_1 \\ \mu_1 \mid pc \vdash M \Downarrow V \mid \mu_2\end{array}}{\mu \mid pc \vdash L := M \Downarrow (\$\ \mathtt{unit})_{\mathtt{low}} \mid \textit{cons}\ a\ V\ \mu_2}$$

$$\Downarrow\textit{-cast}\ \frac{\begin{array}{c}\mu \mid pc \vdash M \Downarrow V \mid \mu_1 \\ \textbf{Active}\ c \qquad \textbf{Cast}\ V\ ,\ c \rightsquigarrow N \\ \mu_1 \mid pc \vdash N \Downarrow W \mid \mu_2\end{array}}{\mu \mid pc \vdash M \langle c \rangle \Downarrow W \mid \mu_2} \qquad \Downarrow\textit{-if-cast-true}\ \frac{\begin{array}{c}\mu \mid pc \vdash L \Downarrow (\$\ \mathtt{true})_\ell \langle c \rangle \mid \mu_1 \qquad \textbf{Inert}\ c \\ \mu_1 \mid pc \curlyvee \ell \vdash M \Downarrow V \mid \mu_2 \\ \mu_2 \mid pc \vdash V \curlyvee \ell \langle \textit{branch}_c\ A\ c \rangle \Downarrow W \mid \mu_3\end{array}}{\mu \mid pc \vdash \mathtt{if}\ L\ A\ M\ N \Downarrow W \mid \mu_3}$$

$$\Downarrow\textit{-if-cast-false}\ \frac{\begin{array}{c}\mu \mid pc \vdash L \Downarrow (\$\ \mathtt{false})_\ell \langle c \rangle \mid \mu_1 \qquad \textbf{Inert}\ c \\ \mu_1 \mid pc \curlyvee \ell \vdash N \Downarrow V \mid \mu_2 \\ \mu_2 \mid pc \vdash V \curlyvee \ell \langle \textit{branch}_c\ A\ c \rangle \Downarrow W \mid \mu_3\end{array}}{\mu \mid pc \vdash \mathtt{if}\ L\ A\ M\ N \Downarrow W \mid \mu_3}$$

$$\Downarrow\textit{-fun-cast}\ \frac{\begin{array}{c}\mu \mid pc \vdash L \Downarrow V \langle c \rangle \mid \mu_1 \qquad \textbf{Inert}\ c \\ \mu_1 \mid pc \vdash M \Downarrow W \mid \mu_2 \\ \mu_2 \mid pc \vdash \textit{elim-fun-proxy}\ V\ W\ c\ pc \Downarrow V' \mid \mu_3\end{array}}{\mu \mid pc \vdash L\ M \Downarrow V' \mid \mu_3} \qquad \Downarrow\textit{-deref-cast}\ \frac{\begin{array}{c}\mu \mid pc \vdash M \Downarrow V \langle c \rangle \mid \mu_1 \qquad \textbf{Inert}\ c \\ \mu_1 \mid pc \vdash\ !\ V \langle \textit{out}_c\ c \rangle \Downarrow W \mid \mu_2\end{array}}{\mu \mid pc \vdash\ !\ M \Downarrow W \mid \mu_2}$$

$$\Downarrow\textit{-assign?-cast}\ \frac{\begin{array}{c}\mu \mid pc \vdash L \Downarrow V \langle c \rangle \mid \mu_1 \qquad \textbf{Inert}\ c \\ \mu_1 \mid pc \vdash \textit{elim-ref-proxy}\ V\ M\ c\ \text{-}{:=^?}\text{-} \Downarrow W \mid \mu_2\end{array}}{\mu \mid pc \vdash L :=^? M \Downarrow W \mid \mu_2}$$

$$\Downarrow\textit{-assign-cast}\ \frac{\begin{array}{c}\mu \mid pc \vdash L \Downarrow V \langle c \rangle \mid \mu_1 \qquad \textbf{Inert}\ c \\ \mu_1 \mid pc \vdash \textit{elim-ref-proxy}\ V\ M\ c\ \text{-}{:=}\text{-} \Downarrow W \mid \mu_2\end{array}}{\mu \mid pc \vdash L := M \Downarrow W \mid \mu_2}$$

Fig. 15. Big-step operational semantics of $\lambda_{\mathsf{SEC}}^{\Rightarrow}$

$$
\begin{array}{rcl}
\text{frames } F & ::= & \square\, M \mid V\, \square \\
& \mid & \texttt{if}\ \square\ A\ M\ N \mid \texttt{let}\ x = \square\ \texttt{in}\ N \\
& \mid & \texttt{ref}^{\checkmark}\ \ell\ \square \mid\ !\ \square \\
& \mid & \square\ \texttt{:=}^{\checkmark}\ M \mid V\ \texttt{:=}^{\checkmark}\ \square \\
& \mid & \square\ \texttt{:=}^{?}\ M \\
& \mid & \square\,\langle c \rangle \mid \texttt{cast}_{\texttt{pc}}\ g\ \square
\end{array}
$$

$$\boxed{\mathit{plug} : \mathit{Term} \longrightarrow \mathit{Frame} \longrightarrow \mathit{Term}}$$

$$
\begin{array}{rcl}
\mathit{plug}\ L\ (\square\ M) & = & L\ M \\
\mathit{plug}\ M (V\ \square) & = & V\ M \\
\mathit{plug}\ L\ (\texttt{if}\ \square\ A\ M\ N) & = & \texttt{if}\ L\ A\ M\ N \\
\mathit{plug}\ M\ (\texttt{let}\ x = \square\ \texttt{in}\ N) & = & \texttt{let}\ x = M\ \texttt{in}\ N \\
\mathit{plug}\ M\ (\texttt{ref}^{\checkmark}\ \ell\ \square) & = & \texttt{ref}^{\checkmark}\ \ell\ M \\
\mathit{plug}\ M\ (!\ \square) & = & !\ M \\
\mathit{plug}\ L\ (\square\ \texttt{:=}^{\checkmark}\ M) & = & L\ \texttt{:=}^{\checkmark}\ M \\
\mathit{plug}\ M\ (V\ \texttt{:=}^{\checkmark}\ \square) & = & V\ \texttt{:=}^{\checkmark}\ M \\
\mathit{plug}\ L\ (\square\ \texttt{:=}^{?}\ M) & = & L\ \texttt{:=}^{?}\ M \\
\mathit{plug}\ M\ (\square\,\langle c \rangle) & = & M\,\langle c \rangle \\
\mathit{plug}\ M\ (\texttt{cast}_{\texttt{pc}}\ g\ \square) & = & \texttt{cast}_{\texttt{pc}}\ g\ M
\end{array}
$$

Fig. 16. Evaluation frames and plug

$$\epsilon : \mathit{Term} \longrightarrow \mathit{Term},\ \epsilon : \mathit{HalfHeap} \longrightarrow \mathit{HalfHeap}, \text{ and } \epsilon : \mathit{Heap} \longrightarrow \mathit{HalfHeap}$$

$$\epsilon\ (\mathtt{addr}\ n_{\hat{\ell}})_{\ell} = \begin{cases} (\mathtt{addr}\ n_{\mathtt{low}})_{\mathtt{low}}, \text{ if } \hat{\ell} = \mathtt{low} \text{ and } \ell = \mathtt{low} \\ \bullet, \text{ if } \hat{\ell} = \mathtt{high} \text{ or } \ell = \mathtt{high} \end{cases} \tag{21}$$

$$\epsilon\ (\$\ k)_{\ell} = \begin{cases} (\$\ k)_{\mathtt{low}}, \text{ if } \ell = \mathtt{low} \\ \bullet, \text{ if } \ell = \mathtt{high} \end{cases} \tag{22}$$

$$\epsilon\ (\lambda^{pc} x{:}A.\ N)_{\ell} = \begin{cases} (\lambda^{pc} x{:}A.\ \epsilon\ N)_{\mathtt{low}}, \text{ if } \ell = \mathtt{low} \\ \bullet, \text{ if } \ell = \mathtt{high} \end{cases} \tag{23}$$

$$\epsilon\ x = x$$

$$\epsilon\ (L\ M) = (\epsilon\ L)\ (\epsilon\ M)$$

$$\epsilon\ (\mathtt{if}\ L\ A\ M\ N) = \mathtt{if}\ (\epsilon\ L)\ A\ (\epsilon\ M)\ (\epsilon\ N)$$

$$\epsilon\ (\mathtt{let}\ x = M\ \mathtt{in}\ N) = \mathtt{let}\ x = (\epsilon\ M)\ \mathtt{in}\ (\epsilon\ N)$$

$$\epsilon\ (\mathtt{ref}\ \ell\ M) = \mathtt{ref}\ \ell\ (\epsilon\ M)$$

$$\epsilon\ (\mathtt{ref}^{?}\ \ell\ M) = \mathtt{ref}^{?}\ \ell\ (\epsilon\ M)$$

$$\epsilon\ (\mathtt{ref}^{\checkmark}\ \ell\ M) = \mathtt{ref}^{\checkmark}\ \ell\ (\epsilon\ M)$$

$$\epsilon\ (!\ M) = !\ (\epsilon\ M)$$

$$\epsilon\ (L := M) = (\epsilon\ L) := (\epsilon\ M)$$

$$\epsilon\ (L :=^{?} M) = (\epsilon\ L) :=^{?} (\epsilon\ M)$$

$$\epsilon\ (L :=^{\checkmark} M) = (\epsilon\ L) :=^{\checkmark} (\epsilon\ M)$$

$$\epsilon\ (M \langle c \rangle) = \epsilon\ M \tag{24}$$

$$\epsilon\ (\mathtt{cast}_{\mathtt{pc}}\ g\ M) = \epsilon\ M \tag{25}$$

$$\epsilon\ \text{-} = \bullet$$

$$\epsilon\ [] = [] \tag{26}$$

$$\epsilon\ (\langle n, V \rangle :: \mu_{\mathtt{low}}) = \langle n, \epsilon\ V \rangle :: (\epsilon\ \mu_{\mathtt{low}}) \tag{27}$$

$$\epsilon\ \langle \mu_{\mathtt{low}}, \mu_{\mathtt{high}} \rangle = \epsilon\ \mu_{\mathtt{low}} \tag{28}$$

Fig. 17. Erasure of $\lambda^{\Rightarrow}_{\mathsf{SEC}}$ terms and the heap

$$\boxed{\mu \mid pc \vdash M \Downarrow_\epsilon V \mid \mu'}$$

$$\Downarrow_\epsilon\text{-}\mathit{val}\ \frac{}{\mu \mid pc \vdash V \Downarrow_\epsilon V \mid \mu} \qquad \Downarrow_\epsilon\text{-}\mathit{app}\ \frac{\begin{array}{c}\mu \mid pc \vdash L \Downarrow_\epsilon (\lambda^{pc'} x{:}A.\ N)_{\texttt{low}} \mid \mu_1 \\ \mu_1 \mid pc \vdash M \Downarrow_\epsilon V \mid \mu_2 \\ \mu_2 \mid pc \vdash N[x := V] \Downarrow_\epsilon W \mid \mu_3\end{array}}{\mu \mid pc \vdash L\ M \Downarrow_\epsilon W \mid \mu_3}$$

$$\Downarrow_\epsilon\text{-}\mathit{app}\text{-}\bullet\ \frac{\begin{array}{c}\mu \mid pc \vdash L \Downarrow_\epsilon \bullet \mid \mu_1 \\ \mu_1 \mid pc \vdash M \Downarrow_\epsilon V \mid \mu_2\end{array}}{\mu \mid pc \vdash L\ M \Downarrow_\epsilon \bullet \mid \mu_2}$$

$$\Downarrow_\epsilon\text{-}\mathit{if}\text{-}\mathit{true}\ \frac{\begin{array}{c}\mu \mid pc \vdash L \Downarrow_\epsilon (\$\ \texttt{true})_{\texttt{low}} \mid \mu_1 \\ \mu_1 \mid pc \vdash M \Downarrow_\epsilon V \mid \mu_2\end{array}}{\mu \mid pc \vdash \texttt{if}\ L\ A\ M\ N \Downarrow_\epsilon V \mid \mu_2} \qquad \Downarrow_\epsilon\text{-}\mathit{if}\text{-}\mathit{false}\ \frac{\begin{array}{c}\mu \mid pc \vdash L \Downarrow_\epsilon (\$\ \texttt{false})_{\texttt{low}} \mid \mu_1 \\ \mu_1 \mid pc \vdash N \Downarrow_\epsilon V \mid \mu_2\end{array}}{\mu \mid pc \vdash \texttt{if}\ L\ A\ M\ N \Downarrow_\epsilon V \mid \mu_2}$$

$$\Downarrow_\epsilon\text{-}\mathit{if}\text{-}\bullet\ \frac{\mu \mid pc \vdash L \Downarrow_\epsilon \bullet \mid \mu_1}{\mu \mid pc \vdash \texttt{if}\ L\ A\ M\ N \Downarrow_\epsilon \bullet \mid \mu_1} \qquad \Downarrow_\epsilon\text{-}\mathit{let}\ \frac{\begin{array}{c}\mu \mid pc \vdash M \Downarrow_\epsilon V \mid \mu_1 \\ \mu_1 \mid pc \vdash N[x := V] \Downarrow_\epsilon W \mid \mu_2\end{array}}{\mu \mid pc \vdash \texttt{let}\ x = M\ \texttt{in}\ N \Downarrow_\epsilon W \mid \mu_2}$$

$$\Downarrow_\epsilon\text{-}\mathit{deref}\ \frac{\begin{array}{c}\mu \mid pc \vdash M \Downarrow_\epsilon (\texttt{addr}\ n_{\texttt{low}})_{\texttt{low}} \mid \mu_1 \\ \mathit{lookup}\ \mu_1\ n = V\end{array}}{\mu \mid pc \vdash\ !\ M \Downarrow_\epsilon V \mid \mu_1} \qquad \Downarrow_\epsilon\text{-}\mathit{deref}\text{-}\bullet\ \frac{\mu \mid pc \vdash M \Downarrow_\epsilon \bullet \mid \mu_1}{\mu \mid pc \vdash\ !\ M \Downarrow_\epsilon \bullet \mid \mu_1}$$

$$\Downarrow_\epsilon\text{-}\mathit{ref?}\ \frac{\mu \mid pc \vdash M \Downarrow_\epsilon V \mid \mu_1 \qquad n = \mathit{length}\ \mu_1 \qquad pc \preccurlyeq \texttt{low}}{\mu \mid pc \vdash \texttt{ref}^?\ \texttt{low}\ M \Downarrow_\epsilon (\texttt{addr}\ n_{\texttt{low}})_{\texttt{low}} \mid \langle n, V\rangle :: \mu_1} \qquad \Downarrow_\epsilon\text{-}\mathit{ref?}\text{-}\bullet\ \frac{\mu \mid pc \vdash M \Downarrow_\epsilon V \mid \mu_1}{\mu \mid pc \vdash \texttt{ref}^?\ \texttt{high}\ M \Downarrow_\epsilon \bullet \mid \mu_1}$$

$$\Downarrow_\epsilon\text{-}\mathit{ref}\ \frac{\mu \mid pc \vdash M \Downarrow_\epsilon V \mid \mu_1 \qquad n = \mathit{length}\ \mu_1}{\mu \mid pc \vdash \texttt{ref}\ \texttt{low}\ M \Downarrow_\epsilon (\texttt{addr}\ n_{\texttt{low}})_{\texttt{low}} \mid \langle n, V\rangle :: \mu_1} \qquad \Downarrow_\epsilon\text{-}\mathit{ref}\text{-}\bullet\ \frac{\mu \mid pc \vdash M \Downarrow_\epsilon V \mid \mu_1}{\mu \mid pc \vdash \texttt{ref}\ \texttt{high}\ M \Downarrow_\epsilon \bullet \mid \mu_1}$$

$$\Downarrow_\epsilon\text{-}\mathit{assign?}\ \frac{\begin{array}{c}\mu \mid pc \vdash L \Downarrow_\epsilon (\texttt{addr}\ n_{\texttt{low}})_{\texttt{low}} \mid \mu_1 \\ \mu_1 \mid pc \vdash M \Downarrow_\epsilon V \mid \mu_2 \qquad pc \preccurlyeq \texttt{low}\end{array}}{\mu \mid pc \vdash L :=^? M \Downarrow_\epsilon (\$\ \texttt{unit})_{\texttt{low}} \mid \langle n, V\rangle :: \mu_2} \qquad \Downarrow_\epsilon\text{-}\mathit{assign?}\text{-}\bullet\ \frac{\begin{array}{c}\mu \mid pc \vdash L \Downarrow_\epsilon \bullet \mid \mu_1 \\ \mu_1 \mid pc \vdash M \Downarrow_\epsilon V \mid \mu_2\end{array}}{\mu \mid pc \vdash L :=^? M \Downarrow_\epsilon (\$\ \texttt{unit})_{\texttt{low}} \mid \mu_2}$$

$$\Downarrow_\epsilon\text{-}\mathit{assign}\ \frac{\begin{array}{c}\mu \mid pc \vdash L \Downarrow_\epsilon (\texttt{addr}\ n_{\texttt{low}})_{\texttt{low}} \mid \mu_1 \\ \mu_1 \mid pc \vdash M \Downarrow_\epsilon V \mid \mu_2\end{array}}{\mu \mid pc \vdash L := M \Downarrow_\epsilon (\$\ \texttt{unit})_{\texttt{low}} \mid \langle n, V\rangle :: \mu_2} \qquad \Downarrow_\epsilon\text{-}\mathit{assign}\text{-}\bullet\ \frac{\begin{array}{c}\mu \mid pc \vdash L \Downarrow_\epsilon \bullet \mid \mu_1 \\ \mu_1 \mid pc \vdash M \Downarrow_\epsilon V \mid \mu_2\end{array}}{\mu \mid pc \vdash L := M \Downarrow_\epsilon (\$\ \texttt{unit})_{\texttt{low}} \mid \mu_2}$$

Fig. 18. Big-step operational semantics of erased $\lambda_{\texttt{SEC}}^{\Rightarrow}$